\documentclass[twocolumn,showpacs,amsmath,amssymb,aps,prd,floats]{revtex4-1}
\usepackage{graphicx}% Include figure files
\usepackage{dcolumn}% Align table columns on decimal point
\usepackage{bm}% bold math
\usepackage{epsfig}
\usepackage{xspace}

\def\apj{The Astrophysical Journal}
\def\apjl{The Astrophysical Journal Letter}

\def\prd{Physical Review D}

\def\araa{Annu. Rev. Astron. Astrophys}
\def\aap{Astronomy and Astrophysics}

\def\mnras{Mon. Not. R. Astron. Soc.}
\def\jcap{JCAP}
\def\apss{Astrophysics and Space Science}

\begin{document}

\title{A hadronuclear interpretation of a high-energy neutrino event coincident with a blazar flare}

\author{Ruo-Yu Liu$^1$}\email{ruoyu.liu@desy.de}
\author{Kai Wang$^{2,4}$}
\author{Rui Xue$^3$}
\author{Andrew M. Taylor$^1$}
\author{Xiang-Yu Wang$^3$}
\author{Zhuo Li$^{2,4}$}
\author{Huirong Yan$^{1,5}$}
\affiliation{$^1$Deutsches Elektronen Synchrotron (DESY), Platanenallee 6, D-15738 Zeuthen, Germany;\\
$^2$Department of Astronomy, School of Physics, Peking University, Beijing 100871, China;\\
$^3$School of Astronomy and Space Science, Nanjing University, Nanjing, 210093, China;\\
$^4$Kavli Institute for Astronomy and Astrophysics, Peking University, Beijing 100871, China;\\
$^5$Institut f\"ur Physik und Astronomie, Universit\"at Potsdam, D-14476 Potsdam, Germany}

\begin{abstract}
Although many high-energy neutrinos detected by the IceCube telescope are believed to have an extraterrestrial origin, their astrophysical sources remain a mystery. Recently, an unprecedented discovery of a high-energy muon neutrino event coincident with a multiwavelength flare from a blazar, TXS~0506+056, shed some light on the origin of the neutrinos. It is usually believed that a blazar is produced by a relativistic jet launched from an accreting supermassive black hole (SMBH). Here we show that the high-energy neutrino event can be interpreted by the inelastic hadronuclear interactions between the accelerated cosmic-ray protons in the relativistic jet and the dense gas clouds in the vicinity of the SMBH. Such a scenario only requires a moderate proton power in the jet, which could be much smaller than that required in the conventional hadronic model which instead calls upon the photomeson process. Meanwhile, the flux of the multiwavelength flare from the optical to gamma-ray band can be well explained by invoking a second radiation zone in the jet at a larger distance to the SMBH. In our model, the neutrino emission lasts a shorter time than the multiwavelength flare so the neutrino event is not necessarily correlated with the flare but it is probably accompanied by a spectrum hardening above a few GeV. 
\end{abstract}

%\pacs{95.85Ry, 98.70Qy, 98.70Sa}
%\date{\today}
\maketitle
\section{Introduction}
Detection of extraterrestrial high-energy neutrinos opens a new era of neutrino astronomy\citep{IC13_sci}. The approximate isotropic distribution of these neutrino events in the sky suggests a large fraction comes from extragalactic sources. It is commonly accepted that high-energy neutrinos are produced in the hadronic interactions of high-energy cosmic rays with matter or with photon fields inside the sources, in which charged pions are generated and give birth to neutrinos when they decay. Various extragalactic astrophysical objects, such as starburst galaxies (e.g.,\citep{Liu14, Tamborra14, Chang15}), tidal disruption events (e.g.,\citep{Wang16, Lunardini17, Senno17}), active galactic nuclei (AGN) (e.g.,\citep{Stecker13, Murase14, Tavecchio15, Petropoulou15, Padovani15}), have been investigated as the possible neutrino sources. 
%
%Among various frequently discussed extragalactic sources such as starburst galaxies (ref), tidal disruption events (ref), gamma-ray bursts (ref), active galactic nuclei including blazars (ref) and radio galaxies (ref), the BL Lacertae (BL Lac) objects which is a species of blazars, have a unique advantage from the point of view of multi-messenger study. This is because that gamma-ray photons are expected to be produced in the same process for neutrino production ,  The BL Lac objects have been found to be the main contributor of the extragalactic gamma-ray sky.
%
Recently, IceCube detected a very-high-energy muon neutrino event IC-170922A on 22 September 2017 which was identified by the  Extremely High Energy (EHE) track event selection \cite{IC17_GCN}. The energy of the neutrino event is estimated to be between 200\,TeV and 7.5\,PeV at 90\%C.L. with the most probable energy to be $\sim 300$\,TeV, by assuming a power-law neutrino spectrum with an index of $-2$ \citep{IC18_TXS}. Coincidently, The \emph{Fermi} Large Area Telescope ({\it Fermi}-LAT) reported that a blazar, or more specifically, a BL Lac object TXS~0506+056 at redshift $z=0.3365$ \citep{Paiano18} is located inside the event error region of $1^\circ$,  {with an increase of the $0.1-300$GeV flux by a factor of 6 during 2018 September 15--27 compared to the 3FGL flux \cite{Fermi17_atel}}. The follow-up observations on this object by various telescopes in various wavelengths also returned positive detections, including a significant detection by MAGIC telescopes at $>100\,$GeV \citep{MAGIC17_atel}, X-ray emissions by \emph{Swift}/XRT and \emph{NuSTAR} \citep{X17_atel}, optical emissions by the ASAS-SN survey and various telescopes \citep{ASASSN17_atel, *Subaru17_atel, *Kanata17_atel, *VLT17_atel}, as well as emission in radio band by VLA\citep{VLA17_atel}. The chance coincidence of the high-energy neutrino event with the multiwavelength flare is disfavoured at the $3\sigma$ level \citep{IC18_TXS}, suggesting the BL Lac object TXS~0505+056 may be counterpart of the neutrino event and hence a cosmic ray (CR) source.

BL Lac objects are regarded as a species of AGN in the unification schemes, with a relativistic jet pointing closely to the observer. {The SMBH that supplies the jet is usually found to be surrounded by partially ionised high-density clouds emitting broad lines at a distance of $d_{\rm BLR}=0.001-0.1\,$pc to the SMBH, and hence the region is also known as the broad line region (BLR). It is usually believed that the BLR reprocesses a fraction of the luminosity of the SMBH accretion disk into its own emission.} If the launched jet extracts a lot of energy from the SMBH, the disk emission is relatively weak in a picture of jet-disk symbiosis \citep{Donea03}, leading to a low luminosity of the BLR. The nondetection of the BLR emission from TXS~0506+056 then could be due to a low BLR luminosity outshone by the bright nonthermal emission from the jet, {similar to the concept of the ``masquerading'' BL Lac as suggested in Ref.\cite{Giommi13}}. Thus, we can still assume the presence of high-density BLR clouds in the vicinity of the SMBH for TXS~0506+056 \footnote{During the review period of this paper, we noticed a new posted paper on arXiv suggesting that TXS~0506+056 is a ``masquerading'' BL Lac and an intrinsically flat-spectrum radio quasar with hidden BLR, see P.~Padovani, F.~Oikonomou, M.~Petropoulou, P.~Giommi, and E.~Resconi, Mon. Not. R. Astron. Soc. {\bf 484}, 104 (2019).}. Actually, {possible indications of BLR emission has been found} in other BL Lac objects \citep[e.g.][]{Vermeulen95, Corbett00, Sbarufatti06, Capetti10, Nilsson10, Landoni12, Landoni18}, typically with a luminosity of $10^{40}-10^{42}\,\rm erg~s^{-1}$). The BLR clouds orbit the SMBH and naturally provide targets for inelastic hadronuclear interactions or proton-proton (hereafter, $pp$) collisions once they enter the jet\citep{Dar97, Araudo10}.

In this work, we will study the neutrino production in the BLR via interactions between CR protons accelerated in the jet and clouds that enter the jet. We will show that a sufficient neutrino production rate can be expected in this scenario to explain the IceCube detection with the jet's proton power being still smaller than the Eddington luminosity of the SMBH. The multiwavelength flux can be reproduced simultaneously by invoking a second radiation zone. The rest part of this paper is organized as follows: we describe the physical picture of our model in Section II. We perform calculation and show the results in Section III. The discussion and the conclusion are presented in Section IV and Section V, respectively.

\section{General picture of the Model}

\begin{figure*}[t]
\centering
\includegraphics[width=1\textwidth]{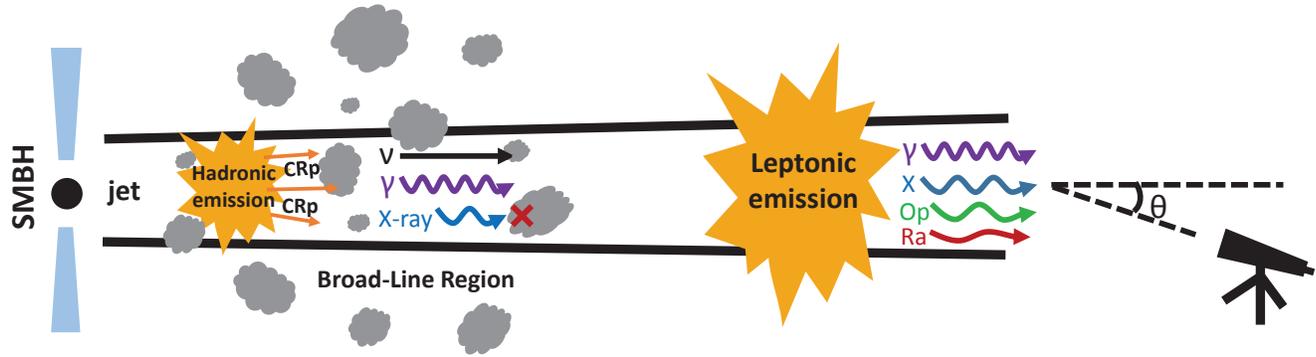}
\caption{A sketch illustration (not to scale) for our model. See text for more details.}
\end{figure*}

The atomic density of a single BLR cloud is $n_c\sim 10^9-10^{11}\,\rm cm^{-3}$, with a size of the cloud $r_c\sim 10^{13}-10^{14}\,$cm \citep{Peterson06, Netzer15}. The typical number of individual BLR clouds is estimated to be $\sim 10^7$. There may also exist diffuse inter-cloud medium of lower-density in the BLR \citep{Peterson06}. The typical mass of BLR in a typical AGN, estimated from line emissions, is about $M_{\rm BLR}\sim 1\,M_\odot$, but there probably exists much more gas emitting less efficiently and hence the total mass of gas in the BLR may be much higher, up to $10^3-10^4M_\odot$ in the extreme case \cite{Baldwin03}. This provides a high gas column density $N_{\rm H}\simeq 10^{24}(M_{\rm BLR}/10\,M_\odot)(d_{\rm BLR}/3\times 10^{16}\,\rm cm)^{-2}\,\rm cm^{-2}$ for neutrino production. Considering the possbility that heightened activity of the SMBH enhances the jet moving with an average bulk Lorentz factor $\Gamma_j$ (or a bulk velocity of $v_j=c\sqrt{1-\Gamma_j^{-2}}$). Some dissipation processes, such as internal collisions between different parts of the jet due to the velocity inhomogeneity, or via the internal-collision-induced magnetic reconnection and turbulence \cite{Zhang11} mechanism, may occur inside or close to the BLR. If the jet loads a certain amount of protons, a fraction of the protons can be accelerated to relativistic energies and interact with clouds in the BLR via the $pp$ collision. Assuming these CR protons will move the jet flow, the BLR clouds provide an interaction efficiency of $f_{pp}=\kappa\sigma_{pp}N_{\rm H}=0.03(N_{\rm H}/10^{24}\rm \,cm^{-2})$ where $\kappa\simeq 0.5$ is the inelasticity of the interaction and $\sigma_{pp}\simeq 50\,$mb is the cross section of the $pp$ collision. From this, one can see that due to the high gas density in the BLR, a proton can lose a considerable fraction of energy in the $pp$ collision.

High-energy electron/positron pairs and gamma-ray photons are also generated in the $pp$ collisions along with neutrinos. Unlike neutrinos, high-energy electrons/positrons and gamma-ray photons can initiate electromagnetic (EM) cascades in the BLR, by interacting with photon fields, magnetic fields and matter in the system via various mechanisms: for relativistic electron/positron, there are mainly three radiation processes, namely, synchrotron radiation in the magnetic field, inverse Compton radiation in the photon field and bremsstrahlung radiation in high-density gas, giving rise to multiwavelength emission; for gamma rays, the main interaction is the $\gamma\gamma$ annihilation with the background photon field in the BLR. An electron/positron pair will be generated in each $\gamma\gamma$ annihilation. For simplicity, we assume a homogeneous distribution of the photon density inside the BLR. The photon spectrum is assumed to be a grey body distribution with a dilution factor $c_{\rm BLR}$ which is obtained by $L_{\rm BLR}R_{\rm BLR}/c=c_{\rm BLR}aT_{\rm BLR}^4R_{\rm BLR}^3$. Here $a$ is the radiation density constant and the temperature $T_{\rm BLR}$ is assumed to be $22000\,K$ so that after multiplying the Boltzmann constant $k$ we have $kT_{\rm BLR}=1.9\,$eV which is the energy of the H$\alpha$ emission line. The intrinsic BLR luminosity of an AGN is usually comparable or several times larger than its narrow line luminosity \citep{Stern12}, while the latter one of TXS~0506+056 is found to be a few times $10^{41}\rm \,erg~s^{-1}$\citep{Paiano18}. For reference, we assume an intrinsic BLR luminosity $L_{\rm BLR}\sim 3\times 10^{41}\rm erg~s^{-1}$, such that the photon number density in the BLR around the peak energy $\varepsilon_p\simeq 2.82kT_{\rm BLR}=5.4\,$eV of the spectrum is $n_{\rm ph}\simeq 10^{10}(L_{\rm BLR}/3\times 10^{41}\,{\rm ergs^{-1}})(R_{\rm BLR}/10^{16}\,{\rm cm})^{-2}(kT_{\rm BLR}/1.9\,\rm eV)^{-1}\rm \,cm^{-3}$. Gamma-ray photons around 100\,GeV will be absorbed by the photon field of the BLR, with an optical depth  $\tau_{\gamma\gamma}\simeq n_{\rm BLR}\sigma_{\gamma\gamma}R_{\rm BLR}\simeq 10$ where $\sigma_{\gamma\gamma}\simeq 10^{-25}\,\rm cm^2$ is approximately the peak cross section of the $\gamma\gamma$ annihilation. The typical energy of electrons/positrons generated by 100\,GeV photons is 50\,GeV. These electrons/positrons will subsequently radiate $\sim 10\,$GeV photons via inverse Compton scattering off the grey body radiation from the BLR with typical energies of a few eV. As a result, the 10\,GeV gamma-ray flux will be enhanced. 
%In addition, a fraction of the 10\,GeV photon can still be absorbed by photons from the high-energy tail of the grey body radiation of the BLR and produce second-generation electron/positron pairs. Such a cascade process will develop in the BLR, but the radiation of higher-generation particles is not important since the gamma-ray opacity is not very large. 
Note that the interaction rate of the photomeson process is roughly three orders of magnitude smaller than that of the $\gamma\gamma$ annihilation with the same target photon field \citep{Dermer12, Murase16}, the photomeson process is henceforth negligible given a $\gamma\gamma$ annihilation opacity of only $\simeq 10$. %Another thing worth noting is that there may also exist a dusty torus extending to a distance of $\sim 0.1-10\,$pc from the SMBH, supplying an infrared photon field. In order not to introduce too many free parameters in our model, we do not consider the inclusion of infrared radiation field, but we show in the Supplement that including them is not expected to change our result.

The synchrotron radiation of electrons/positrons generated in cascades can produce strong UV/X-ray emission. BLR clouds that enter the jet will be fully ionised by the UV/X-ray emission. Due to the high column density of BLR we consider, the ionised electrons will provide a large opacity for optical to X-ray photons by Compton scattering, while the gamma-ray photons escape due to the suppressed cross section (i.e., Klein-Nishina effect). To explain the multiwavelength emission, we invoke a second radiation zone beyond the BLR where the kinetic energy of the jet is dissipated, such as a dissipating ``blob'' which is usually employed to explain the multiwavelength emission of BL Lac objects in many previous models \citep[e.g.][]{Inoue96, Tavecchio98, Costamante02}.     
We ascribe the multiwavelength emission to the synchrotron radiation and synchrotron self-Compton (SSC) of the nonthermal electrons accelerated in the blob. Note that having two (or more) radiation zones may not be unnatural. For example, if the dissipations are produced by internal collisions due to inhomogeneity in the jet speed, multiple collisions can occur at different places and form multiple radiation zones. Actually, we have also seen many bright knots distributing along the jet of radio galaxies (e.g.,\citep{Chartas00,Jester06}). If we observe these sources on the jet axis, we will see the superposition of the emissions from all those knots. 
%{Note that the velocity of blob outside the BLR may be smaller (but still relativistic) since some cloud material may be loaded in the jet when the jet pass through the BLR (see Supplement for discussion)}. 
The key difference between the dissipation in the BLR and outside the BLR is the environment in which the dissipation takes place. {If the dissipation does not take place inside or close to the BLR, then there would be too few target gas in the dissipation region for efficient $pp$ collision and subsequently little neutrino will be produced}. Due to this reason, the neutrino emission is not necessarily expected to be temporally associated with the low-energy emission.

\section{Methods and Results}

\subsection{Hadronic emission in the BLR}
\begin{figure}[htbp]
\centering
\includegraphics[width=0.9\columnwidth]{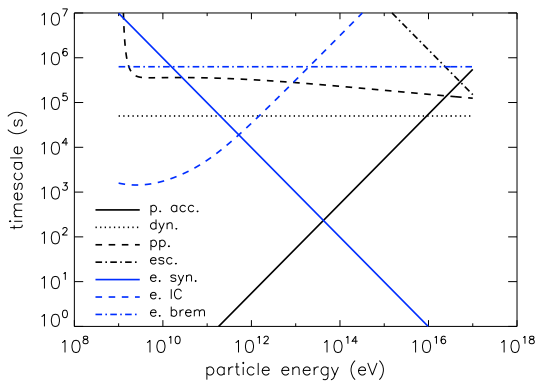}
\caption{Timescales of various processes in the BLR measured in the jet comoving frame. The black dotted, dashed, solid lines represent the dynamical timescale, $pp$ collision timescale and proton acceleration timescale respectively. The blue solid, dashed, dash-dotted line represent the synchrotron cooling timescale, inverse Compton cooling timescale and bremsstrahlung cooling timescale for secondary electrons respectively. Parameters are the same with the ones shown in Table.~\ref{tab:parameters} for $s=2.0$.}\
\label{fig:blr_timescales}
\end{figure}

We denote the total luminosity of nonthermal protons (i.e., the power of accelerated protons) in the jet comoving frame by $L_{p,\rm BLR}'$ (hereafter primed quantities represent the quantities in the jet comoving frame), and assume the differential proton spectrum at injection to be $\dot{N}_p'\propto E_p'^{-s}{\rm exp}(-E_{p,\rm max}')$ in the jet comoving frame, {spanning from the minimum energy 1\,GeV to} the maximum achievable proton energy in the acceleration $E_{p,\rm max}'$. Since the produced neutrino takes about 5\% of the energy of the parent proton, to produce a neutrino of energy $E_\nu$, the proton energy in the jet frame needs to be $E_p'\simeq 20E_\nu/\Gamma_j=10^{15}(E_\nu/10^{15})(\Gamma_j/20)^{-1}\,$eV.
Generally, the proton acceleration timescale can be estimated by $t_{\rm acc}'\simeq 1000\eta (E_p'/10^{15}{\rm eV})(B_j'/0.1\,\rm G)^{-1}\,$s,  where $B_j'$ is the magnetic field in the jet and $\eta\geq 1$ is a prefactor depending on the diffusion of CRs. The acceleration of a proton to this energy is required to be accomplished before the proton loses a significant fraction of their energies or within the dynamical timescale. So, we need to compare the acceleration timescale to the dynamical timescale $t_{\rm dyn}'\simeq 5\times 10^4(d_{\rm BLR}/3\times 10^{16}\,{\rm cm})(\Gamma_j/20)^{-1}\,$s and the energy loss timescale due to $pp$ collision $t_{pp}'\simeq 7\times 10^5(\Gamma_j/20)^{-1}(n_{\rm H}/10^{8}\rm cm^{-3})^{-1}(\sigma_{pp}/60\rm mb)^{-1}\,$s, where $n_{\rm H}\simeq N_{\rm H}/R_{\rm BLR}$ is the average gas density in the BLR. Protons may also escape the BLR which depends on a detailed specification of the geometry, the boundary conditions, and the local turbulence property. 
Regardless of the complexity, the limit of the escape timescale is roughly $R_{\rm BLR}/\Gamma_jc$ (i.e., ballistic escape) which is comparable to the dynamical timescale. Thus, the uncertainty on the escape timescale will not have significant influence on our results. We show relevant timescales in ~\ref{fig:blr_timescales}). The proton spectrum in the BLR can then be estimated by $N_p'=\dot{N}_p't_p'$ where  $t_p'=(t_{pp}'^{-1}+t_{\rm dyn}'^{-1})^{-1}$. We then can obtain the kinetic luminosity of relativistic protons to be $L_{p,k}=\pi R_j^2\Gamma_j^2c\int E_p'N_p'dE_p'/V'\simeq \Gamma_j^2L_{p,\rm BLR}$ where $R_j$ is the transverse radius of the jet and $V'\simeq \pi R_j^2R_{\rm BLR}/\Gamma$ is approximately the volume of the dissipation zone in the BLR region.

Hadronuclear interactions between accelerated protons and atoms of in the BLR clouds produce neutral and charged pions, which eventually decay into gamma-ray photons, electrons/positrons, and neutrinos, i.e.,
\begin{eqnarray*}
p+p&\to& \pi^0\to \gamma+\gamma\\
p+p&\to& \pi^+\to \nu_\mu+\mu^+\to \nu_\mu+e^++\nu_e+\bar{\nu}_\mu\\
p+p&\to& \pi^-\to \bar{\nu}_\mu+\mu^-\to \bar{\nu}_\mu+e^-+\bar{\nu}_e+\nu_\mu
\end{eqnarray*}
The differential spectrum of the secondary particles produced in unit time are calculated following the semianalytic method developed by \citep{Kelner06} (see also \citep{Stecker70, Ervin14}), i.e., 
\begin{equation}
\dot{N}_i'(E_i')\equiv\frac{dN_i'}{dE_i'dt'} = cn_{\rm H}'\int_{E_i}^{\infty}\sigma_{pp}N_p'(E_p')F_i(\frac{E_i'}{E_p'},E_p')\frac{dE_p'}{E_p'}
\end{equation}
where $i$ could be $\gamma$, $e$ or $\nu$, and $F_i$ is the spectrum of the secondary $\gamma$, $e^\pm$ or $\nu$ in a single collision. This description works for $E_p\gtrsim 100\,$GeV, while for $E_p < 100\,$GeV a $\delta$-functional approximation for the energy of produced pions can be used to obtain the secondary spectrum 
\begin{equation}
\begin{split}
\dot{N}_i'(E_i')&=2cn_{\rm H}'\frac{\tilde{n}}{K_\pi}\int_{E_{i, \rm min}'}^{\infty}\sigma_{pp}\left(m_p+\frac{E_\pi'}{K_\pi}\right)\\
&\times \xi_i\frac{dN_p}{dE_p}\left(m_p+\frac{E_\pi'}{K_\pi}\right)\frac{dE_\pi'}{\sqrt{E_\pi'^2-m_\pi^2}}
\end{split}
\end{equation}
where $E_\pi'$ is the energy of pions and the pion rest mass $m_\pi \simeq 135\,$MeV for gamma-ray production and $m_\pi\simeq 140\,$MeV for neutrino production. $E_{i,\rm min}'=E_i'/\zeta_i+\zeta_i m_\pi^2/4E_i'$, with $\zeta_\gamma=1$ for gamma rays, $\zeta_e=1$ for $e^\pm$, (anti-)muon neutrinos and (anti-)electron neutrinos from $\mu^\pm$ decay), and $\zeta_\nu=1-m_\mu^2/m_\pi^2=0.427$ for (anti-)muon neutrino from $\pi^\pm$ decay. $m_\mu\simeq 106\,$MeV is the muon rest mass), $\xi_\gamma=1$, $\xi_\mu=1$, and $\xi_e=\frac{35}{16}[1-(\frac{E_e'}{E_{e,\rm max}'})^2]^3$ where $E_{e,\rm max}'=(E_\pi'+\sqrt{E_\pi'^2-m_\pi^2})/2$. $K_\pi=0.17$, and $\tilde{n}$ is a free parameter that is determined by the continuity of the flux of the secondary particle at $100\,$GeV. 

The produced neutrinos will escape the radiation zone, but high-energy photons and electrons/positrons will initiated EM cascades. We follow the treatment in (\citep{Boettcher13, Wang18}, also see Supplement for details) to calculate the quasi-steady state cascade emission, since the cooling timescale of electrons is shorter than the dynamical timescale (see blue curves in Fig.~\ref{fig:blr_timescales}). 

\subsection{Ionisation of BLR clouds and Compton opacity for UV/X-ray photons emitted in the BLR}
The electrons generated in the EM cascade radiate UV/X-ray photons via synchrotron radiation which can ionise the BLR clouds. Assuming the clouds are composed of pure hydrogens, the photoionisation cross section can be given by \citep{Draine11}
\begin{equation}
\sigma_{\rm PI}=\left\{
\begin{array}{ll}
\sigma_0\left(\frac{E_\gamma}{I_H} \right)^{-3}, ~~ {\rm for} ~~ I_H<E_\gamma\lesssim 100I_H,\\
\frac{3e^4}{2\pi}\sigma_0\left(\frac{E_\gamma}{I_H} \right)^{-3.5}, ~~{\rm for} ~~ E_\gamma> 100I_H
\end{array}
\right.
\end{equation}
where $I_H=13.6\,$eV is the ionisation energy of atomic hydrogen, $\sigma_0=6.3\times 10^{-18}\,\rm cm^2$ is the cross section at threshold and $e\simeq 2.72$ is the Euler's number. Take $s=2.0$ case as example, we calculate the the photoionisation rate by
\begin{equation}
\zeta_{\rm ion}=\int_{I_H}^{\infty} n_\gamma(E_\gamma)\sigma_{\rm PI}cdE_\gamma\simeq 100 \,\rm s^{-1}
\end{equation}
where $n_\gamma(E_\gamma)$ is the differential photon number density based on the unabsorbed flux emitted by the BLR shown in Fig.~\ref{fig:decompose}. The recombination rate of pure hydrogen gas is given by $\zeta_{\rm rec}=3\times 10^{-6}T^{-1/2}n_H\simeq 6(kT_{\rm BLR}/1.9\,\rm eV)^{-1/2}(n_H/3\times10^8)\,\rm s^{-1}$. If the metallicity of the cloud is not zero, the recombination rate will be further reduced. In addition to photoionisation, some clouds may directly interact with the jet and a shock may be driven in the cloud, and the cloud may also be ionised in this process. Thus, the BLR clouds that entered into the jet will be fully ionised.

Ionised electrons will scatter the photons to other direction from our line of sight. The optical depth is $\tau_{\rm sc}=\sigma_{\rm sc}n_HR_{\rm BLR}\simeq 2$ for $E_\gamma<$MeV where
\begin{equation}
\begin{split}
\sigma_{\rm sc}&=\sigma_{\rm T}\cdot \frac{3}{4}\bigg[\frac{1+x}{x^3}\left\{\frac{2x(1+x)}{1+2x}-\ln(1+2x) \right\}\\
&+\frac{1}{2x}\ln(1+2x)-\frac{1+3x}{(1+2x)^2}\bigg] ,
\end{split}
\end{equation}
with $x=E_\gamma/m_ec^2$. We need to multiply a factor of $(1-{\rm exp}(-\tau_{\rm sc}))/\tau_{\rm sc}$ to the obtained flux in the BLR region, and this will reduced the optical to X-ray flux of the cascade emission but gamma-ray will not be influenced since the scattering cross section is suppressed by the Klein-Nishina effect. Note that such a opacity for X-ray in the BLR is also needed in order not to overshoot the observed flux.

\subsection{Leptonic emission}
Given the Compton opacity of the BLR, to explain the multiwavelength emission, we invoke a second (or more) radiation zone beyond the BLR where the kinetic energy of the jet is dissipated, such as a dissipating ``blob'' which is usually employed to explain the multiwavelength emission of BL Lac objects in many previous models \citep[e.g.][]{Inoue96, Tavecchio98, Costamante02}. Since dissipations take place outside the BLR, there will not be $pp$ collision even if CR protons are accelerated in the blob and hence leptonic emission of accelerated electrons will dominate. 

We assume relativistic electrons are injected in the blob with a luminosity $L_e'$. To reproduce the observed flux in optical to soft X-ray band, we employ a broken power-law function for the electron injection spectrum, with a broken energy $E_{e,b}'$ and spectral index $s_1$ and $s_2$ below and above the break, respectively, i.e., 
\begin{equation}
\dot{N}_{e,\rm blob}'\propto \left\{
\begin{array}{ll}
 \left( \frac{E_e'}{E_{e,b}'}\right)^{-s_1}, ~~ E_{e,0}'\leq E_e'<E_{e,b}'\\
 \left( \frac{E_e'}{E_{e,b}'}\right)^{-s_2}, ~~ E_e'\geq E_{e,b}'
\end{array}
\right.
\end{equation}
with $E_{e,0}'$ being the minimum energy of the injected electron. Similar to the case of protons in the BLR, we can obtain the normalisation of the electron injection spectrum by $\int E_e'\dot{N}_{e,\rm blob}'dE_e'=L_e'$. The total electron spectrum in the blob comoving frame is 
then $N_{e,\rm blob}'=\dot{N}_{e, \rm blob}'t_e'$ where $t_e'=\left(t_{c, \rm blob}'^{-1}, t_{\rm dyn, blob}'^{-1}\right)^{-1}$, representing the electron cooling timescale, and 
the dynamical timescale or the adiabatic expansion timescale of the blob $t_{\rm dyn, blob}'=R_{\rm blob}'/c$ respectively. Electron cools due to the synchrotron radiation and the synchrotron self-Compton scattering (SSC), so we have $t_{c,\rm blob}'=\frac{3m_e^2c^3}{4\sigma_T}\frac{E_e'}{u_{B,\rm blob}'+u_{\rm syn}'}$ where $u_{B,\rm blob}'=B_{\rm blob}'^2/8\pi$ is the energy density of magnetic field in the blob with $B_{\rm blob}'$ being a free parameter, and $u_{\rm syn}'$ is the energy density of synchrotron radiation of the electrons in the blob which can be determined from the observed optical flux. The kinetic luminosity of accelerated electron in the jet can then be obtained by
$L_{e,k}=\pi R_{\rm blob}'^2\Gamma_j^2c\int N_{e,\rm blob}'E_e'dE_e'/V_{\rm blob}'$ where $V_{\rm blob}'=4\pi R_{\rm blob}'^3/3$ is the volume of the blob.

{On the other hand, primary electrons will also be accelerated along with protons inside the BLR. We assume that the injection spectrum of electrons in the BLR is the same as that in the blob outside the BLR. The difference is that these electrons mainly radiate via the inverse Compton scattering off the external radiation field (BLR's radiation field), and the adiabatic cooling of electrons is stronger than that in the blob outside the BLR given the size of the emission zone is smaller inside the BLR. The spectrum of the produced external Compton (EC) radiation peaks at $\sim 10$\,GeV, and does not induce electromagnetic cascades.}\\

After we obtain the differential luminosity of both the emissions from the BLR and the blob in the comoving frame, i.e., $L'(E_\gamma')=L_{\rm BLR}'+L_{\rm blob}'$, we can calculate the flux at the Earth by
\begin{equation}
f_\gamma(E_\gamma)=\frac{\delta_D^4L'(E_\gamma')}{4\pi D_L^2}e^{-\tau_{\gamma\gamma}^{\rm EBL}(E_\gamma,z)}
\end{equation}
where the factor $\delta_D^2$ accounts for the beaming effect due to relativistic motion of the jet (blob) while another $\delta_D^2$ considers the Doppler boost of the flux. $D_L=1.77\,$Gpc is the luminosity distance for the redshift $z=0.3365$, while $E_\gamma=\delta_DE_\gamma'/(1+z)$. $\tau_{\gamma\gamma}^{\rm EBL}$ is the optical depth for gamma-ray photons due to the absorption by the extragalactic background light (EBL). Here we employ the EBL model provided by \citep{Finke10}. Note that a pure leptonic model can also give an acceptable fitting to the multiwavelength flux. The hadronic process is considered mainly for the neutrino production.

\begin{figure*}[htbp]
\centering
\includegraphics[width=1.6\columnwidth]{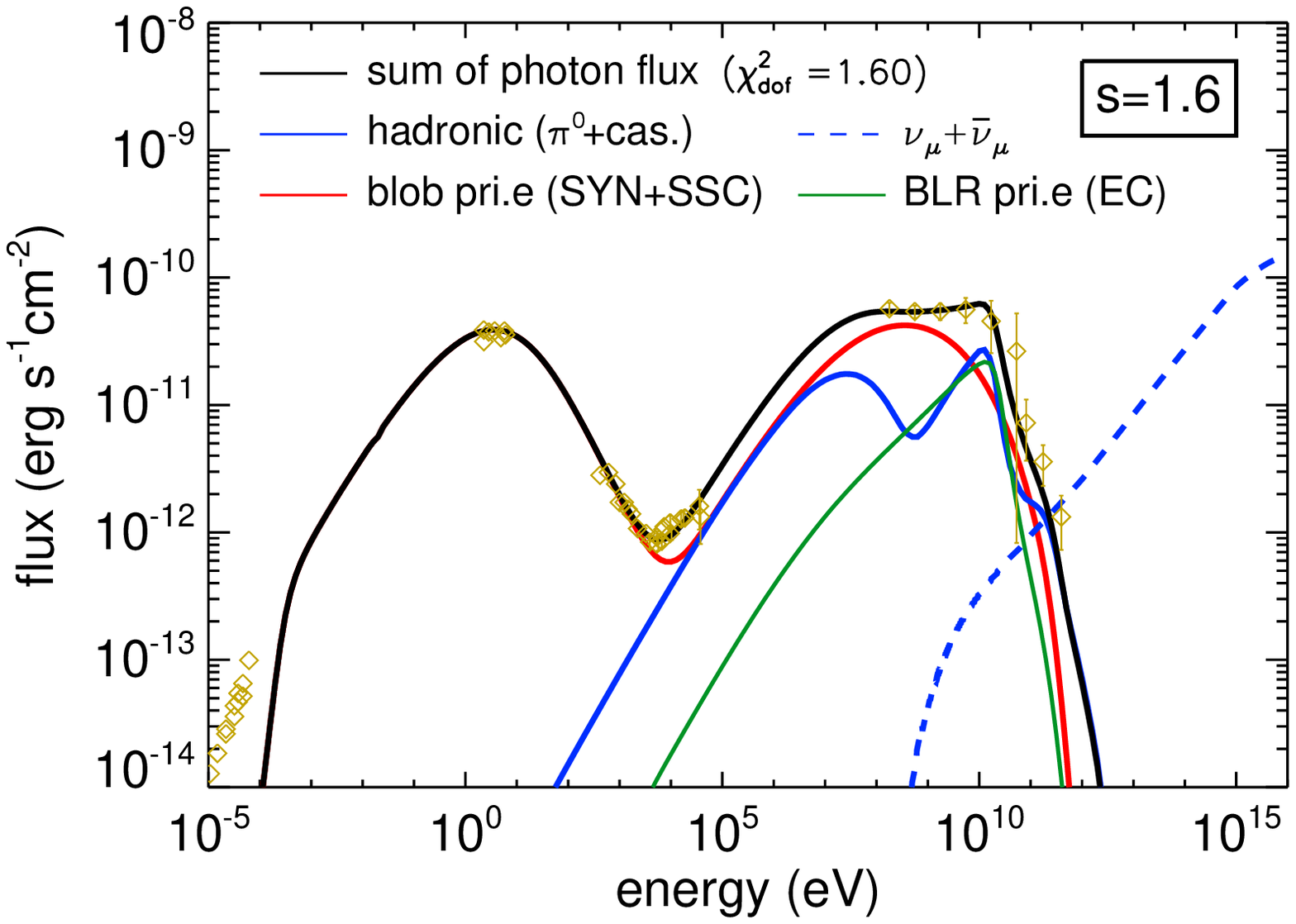}
\includegraphics[width=1.6\columnwidth]{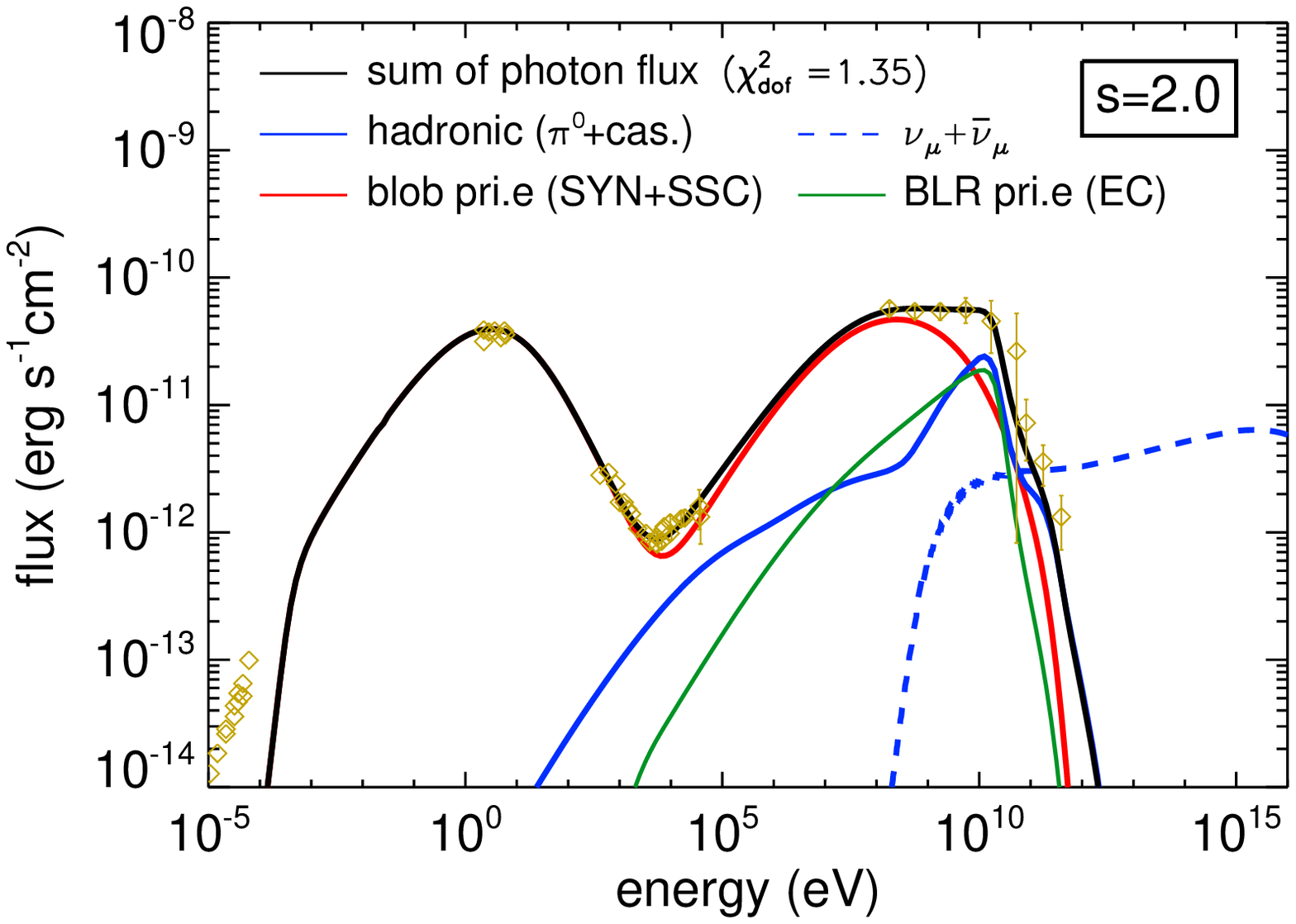}
\caption{Predicted multiwavelength flux and neutrino flux of TXS~0506+056 for $s=1.6$ (upper panel) and $s=2.0$ (lower panel). The red, blue, green curves are the flux from synchrotron and SSC emission of electrons of the blob outside the BLR, and hadronic emission (including pionic emission and EM cascades) and the EC emission of co-accelerated electrons inside the BLR, with the black curves being the sum of them. The synchrotron and SSC flux of co-accelerated electrons inside the BLR is not important (since the cooling due to the EC process is much more important) and thus they are not shown here for the clarity of the figure. The blue dash curves represent the flux of $\nu_{\mu}+\bar{\nu}_{\mu}$ flux assuming a flavor ration of $1:1:1$ after oscillation. The yellow diamonds are data points taken from \citep{IC18_TXS}. To obtain a reasonable reduced chi-square value, we introduce an error of 2\% for each optical data point, which is the typical level of the systematic error \citep{Magic14}. }\label{fig:fitting}
\end{figure*}

\subsection{Results}
We consider two slopes for the accelerated proton spectrum in the BLR, say, $s=1.6$ and $s=2.0$. {The predicted multiwavelength flux and neutrino flux are shown in Fig.~\ref{fig:fitting}, in comparison with the measured multiwavelength data within two weeks of the neutrino detection. Model parameters are given in Table.~\ref{tab:parameters}. We do not optimize the fitting (e.g., minimize the $\chi^2$) noting instead simply that a reasonable reduced $\chi^2$ is obtained. In both two cases, the leptonic emission from the blob outside the BLR makes important contribution to multiwavelength flux, while hadronic emission including the co-accelerated primary electrons in the BLR partly contribute to X-ray and $>10$GeV flux. 
The radio flux can not be fitted in both two cases due to the synchrotron self-absorption by the accelerated electrons. The difficulty of fitting the radio emission has been also found in other BL Lac objects with it being suggested that the radio emission arise from an extended region with a weaker magnetic field (e.g.,\citep{Gao17}). One interesting feature in our model is that the superposition of the SSC emission in the blob, {the EC emission of primary electrons} and hadronic emission in the BLR can reproduce the flat spectrum in $0.1-10\,$GeV as observed by Fermi-LAT, while a pure SSC emission leads to a curved spectral shape. Such a relatively hard spectrum above a few GeV may be an indicator of the neutrino emission. 
%Although the multiwavelength fluxes are similar in both $s=1.6$ and $s=2.0$ cases, the neutrino flux in these two cases are different. For $s=2.0$, the energy of accelerated proton evenly distributes in each energy decade. In contrast, for $s=1.6$, most energy of accelerated proton is above a PeV and the energy below TeV is small, resulting in a higher gamma-ray/neutrino flux above 0.1\,PeV and a lower gamma-ray/neutrino flux below 100\,GeV than that in $s=2.0$ case, given the same proton luminosity. Therefore, a smaller proton luminosity is needed in $s=2.0$ case in order to fit the gamma-ray spectrum and this further lowers the PeV neutrino flux compared to $s=1.6$ case. 
{Based on the effective area of IceCube EHE alerts provided in \citep{IC18_TXS}, which is about $10\,\rm cm^2$ for 200\,TeV neutrino and is roughly proportional to the neutrino energy} in the direction of TXS~0506+056, we find that, {by convolving the predicted neutrino flux with the effective area}, IceCube is expected to detect one muon (or anti-muon) neutrino event in $0.2-7.5\,$PeV in 100 days for $s=1.6$ and in 3.8 years for $s=2.0$, should the SMBH activity lasts such a long period of time.
%On the other hand, since the neutrino emission is not necessarily in temporal association with the multiwavelength flare, we should be cautious to use the multiwavelength flare as an indicator of the neutrino emission. 

\begin{table*}[htpb]
\centering
\caption{Main parameters used in the spectral fittings. {Note that some parameters such as the injection spectral index $s$, temperature of the BLR cloud $T_{\rm BLR}$, distance of the BLR to the black hole $d_{\rm blob}$, bulk Lorentz factor $\Gamma$ (and velocity), viewing angle $\theta$, the BLR column density, the BLR size and etc are not treated as free parameters, but are fixed in the calculation and we do not adjust them to fit the data, whereas paremeters such as the Doppler factor, BLR's mean atom density, dilutoin factor of the BLR emission are not independent parameters. The injection spectrum of primary electrons inside the BLR is assumed to be identical to that in the blob. The number of free parameters in the modeling is 12.}}\label{tab:parameters}
\begin{tabular}{|c|l|c|c|}
\hline\hline
Parameters & Descriptions & \multicolumn{2}{c|}{Values}\\
\hline
$s$        & proton spectral index at injection & $s=1.6$ & $s=2.0$\\        
\hline
$L_{p,k}$ & jet's kinetic luminosity of accelerated protons & $4.5\times 10^{46}\rm erg~s^{-1}$ & $7.7\times 10^{45}\rm erg~s^{-1}$\\
\hline
$L_{\rm BLR}$ & luminosity of the grey body emission of the BLR & $6.4\times 10^{41}\,\rm erg~s^{-1}$ & $3.2\times 10^{41}\,\rm erg~s^{-1}$\\
\hline
$c_{\rm BLR}$ & dilution factor of the grey body emission & $1.2\times 10^{-4}$ & $6\times 10^{-5}$ \\
\hline
$B_{\rm BLR}'$ & magnetic field of the BLR in the jet comoving frame & 0.05\,G & 0.1\,G\\
\hline
$\eta$ & the ratio of the mean free path to Larmor radius of protons in the BLR & 3 & 5\\
\hline
$T_{\rm BLR}$ & temperature of the grey body emission of the BLR & \multicolumn{2}{c|}{1.9\,eV}\\
\hline
$d_{\rm BLR}$ & mean distance of the BLR to the SMBH & \multicolumn{2}{c|}{$3\times 10^{16}\,$cm}\\
\hline
$R_{\rm BLR}$ & size of the BLR & \multicolumn{2}{c|}{$10^{16}$\,cm}\\
\hline
$N_{\rm H}$ & column density of the BLR & \multicolumn{2}{c|}{$10^{24.5}\,\rm cm^{-2}$}\\
\hline
$n_{\rm H}$ & mean atomic density of the BLR & \multicolumn{2}{c|}{$3\times 10^{8}\,\rm cm^{-3}$}\\
\hline
$\Gamma_j$ & bulk Lorentz factor of the jet(blob)$^a$ & \multicolumn{2}{c|}{20}\\
\hline
$\beta_j$ & bulk speed of the jet(blob) in unit of $c$ & \multicolumn{2}{c|}{0.9987}\\
\hline
$\theta$ & viewing angle of the jet(blob) & \multicolumn{2}{c|}{$4^\circ$}\\
\hline
$\delta_D$ & Doppler factor of the jet(blob)$^b$ & \multicolumn{2}{c|}{13.6}\\
\hline
$d_{\rm blob}$ & distance of the blob to the SMBH & \multicolumn{2}{c|}{$3\times 10^{18}$\,cm}\\
\hline
$R_{\rm blob}$ & size of the blob & $10^{16.70}\,$cm & $10^{16.63}\,$cm\\
\hline
$B_{\rm blob}'$ & magnetic field of the BLR in the blob comoving frame & 0.33\,G & 0.48\,G\\
\hline
$L_{e,k}$ & jet's kinetic luminosity of accelerated electrons & $4.2\times 10^{44}\rm erg~s^{-1}$ & $4.0\times 10^{44}\rm erg~s^{-1}$\\
\hline
$E_{e,b}'$ & break energy in the electron spectrum injected to the blob & 6.1\,GeV & 5.1\,GeV\\
\hline
$E_{e,0}'$ & minimum energy of the electron injected to the blob & 0.005\,GeV & 0.005\,GeV\\
\hline
$s_1$  & electron spectral index before the break &  1.55 & 1.55\\
\hline
$s_2$  & electron spectral index after the break & 3.72 & {3.69}\\
\hline
\end{tabular}
\begin{list}{}
\centering
\item{$a$}: the bulk Lorentz factors of the blobs inside the BLR and outside the BLR are not necessarily the same.\\ 
\item{$b$}: $\delta_D=[\Gamma_j(1-\beta_j\cos\theta)]^{-1}$.
\end{list}
\end{table*}

\section{Discussion}
\subsection{Comparison with the photomeson model}
Different from the conventional hadronic model for neutrino production in blazars which considers photomeson process \citep{Stecker91, Mannheim93, Atoyan01, Mucke03}, we ascribe the neutrino production to the $pp$ collision by assuming a high column density gas in the BLR. The efficiency of the hadronic interaction can approach $\sim 10\%$ without introducing too large an internal $\gamma\gamma$ annihilation opacity for gamma rays. {As a consequence, our model results in a moderate proton power of the jet, i.e., $L_p\sim (0.8-5)\times 10^{46}\,\rm erg~s^{-1}$, which is about $(5-30)\%$ of the Eddington luminosity of a SMBH with a mass of $10^9M_\odot$. By contrast, the photomeson model usually leads to a quite low efficiency for neutrino production in order to avoid a large internal $\gamma\gamma$ annihilation opacity for gamma rays and hence has to invoke a huge proton luminosity that far exceeds the Eddington luminosity of the SMBH.} The neutrino spectrum in the $pp$ collision scenario can extend down to GeV energy roughly following the proton spectrum, so in principle we may expect the detection of $<100\,$TeV neutrino from TXS~0506+056 (perhaps relating to the earlier neutrino flare from this source \citep{IC18_TXSnuflare}).

\subsection{Correlation between the neutrino emission and the multiwavelength emission}
In our model, hadronuclear interactions will take place only when the dissipation occurs in the BLR. If the dissipation takes place randomly along the jet axis, there should be more dissipation happening outside the BLR than inside the BLR. As we can see in Fig.~\ref{fig:fitting}, the leptonic emission can solely account for the multiwavelength data, and hadronic emission is responsible only for part of the X-ray and gamma-ray emission. Thus, the neutrino emission is not necessarily correlated with the multiwavelength flare. On the other hand, we expect a spectrum hardening above a few GeV due to the inverse Compton radiation of cascade electrons in the BLR, when the dissipation happens inside the BLR and trigger an efficient neutrino production. For TXS~0506+056, it seems that our prediction is consistent with the Fermi-LAT data within two weeks of the neutrino detection. The spectral hardening above a few GeV may be an indicator of the neutrino emission via the $pp$ collision and can be used to test our model in the future if the  statistics is good enough.

On the other hand, the point-source effective area of IceCube is about 10 times larger than that of the EHE alerts. Our model would predict one event detection in $\sim 10$ days for $s=1.6$ with the point-source effective area, which is comparable to the dynamical timescale of an orbiting BLR cloud crossing the jet (i.e. $t_{c}=(d_{\rm BLR}/3\times10^{16}{\rm cm})^{3/2}(M_{\rm BH}/10^9M_\odot)^{1/2}\simeq 10^6$s, assuming that clouds orbit with Keplerian velocity and jet's width is about 10\% of the jet's length). This is also consistent with a time-dependent analysis using the point-source search, in which a Gaussian time window is employed and no other event around the detection time of IC-170922A was found, resulting in a $\gtrsim 2\sigma$ excess with the time window being centered at 22 September 2017, a duration of 19 days and a spectral index of $1.7\pm0.6$ \citep{IC18_TXSnuflare}. In our interpretation, the neutrino emission lasts a few weeks (unless there are more than one dissipations taking place inside the BLR) and the event IC-170922A is not a lucky detection once a dissipation takes place inside the BLR. 

{Furthermore, interestingly, \citep{IC18_TXSnuflare} also reports an outburst of neutrinos detected from TXS~0506+056 during its quiescent state. Such a discovery favor a hadronuclear origin of the neutrino outburst and may corroborate with our model here, suggesting a gas-rich environment in the vicinity of the supermassive black hole of TXS~0506+056.}

\subsection{Jet-cloud interactions}
In this work, we take an average gas density of the BLR to calculate the $pp$ collision for simplicity, based on the assumed column density and the size of the BLR. In reality, BLR gas probably exist in the form of gas clumps or clouds, as we mentioned in Section ~2. Some clouds may encounter the jet when they orbit the SMBH, and the jet will exert a pressure on the clouds to accelerate the clouds along the jet propagation axis. The encounter also drive shocks expanding into the clouds, and hydrodynamical instabilities can occur leading to the cloud deformation and fragmentation \citep{Araudo10, Pittard10}. According to \citep{Pittard10}, the cloud drag timescale (defined as the time for the relative velocity between the cloud and the ambient flow to to decrease by a factor of $e^{-1}$) and the cloud mixing timescale (defined as the time needed for the mass of the core of the cloud to decrease by a factor of 2) are about one order of magnitude longer than the cloud shocking time, i.e., $\sim 10t_{cc}=10\chi^{1/2}r_c/c\simeq 10^6\,$s given a cloud radius $r_c=10^{14}\,$cm and a density contrast $\chi=3\times 10^3(n_c/10^{11}{\rm cm^{-3}})(L_j/10^{47}{\rm ergs^{-1}})(\Gamma_j/20)(R_j/10^{15}\,\rm cm)^{-2}$ between the cloud and the jet with $L_j$ being the jet's kinetic luminosity and 
$R_j$ being the jet transverse radius. This timescale is comparable to the dynamical timescale $t_{\rm dyn}$ and the time needed by the cloud to cross the jet $t_{c}$. As a result, a considerable fraction of the cloud material may be loaded in the jet and jet is slowed down after passing through the BLR. On the other hand, given the total cloud mass within the jet section $\pi R_j^2m_pN_{\rm H}$ and the mass of a single cloud to be $4\pi r_c^3m_pn_c/3$, we can estimate the total number of clouds in the jet section is $N_c\simeq 25 (\frac{N_H}{10^{24.5}{\rm cm^{-2}}})(\frac{R_j}{10^{15}{\rm cm}})^{-2} (\frac{r_c}{10^{14}{\rm cm}})^{-3}(\frac{n_c}{10^{11}{\rm cm^{-3}}})^{-1}$. The covering fraction of the jet by these clouds is then $N_c(r_c/R_j)^2=0.25$ (note that the covering fraction for hadronic emission is unity since $pp$ collisions take place inside the clouds) if different clouds do not overlap each other along the jet axis. Thus, we speculate the jet will not experience a global deceleration.  Furthermore, even if all the BLR clouds that enter the jet are homogeneously mixed into the jet, the bulk Lorentz factor of the jet decrease to $\Gamma_j/3$ considering conservation of kinetic energy, given the mass of the jet from the base to the BLR is $\simeq L_jd_{\rm BLR}/\Gamma c^3=0.004M_\odot$. For $\Gamma=20$ and a viewing angle of $4^\circ$ as employed in the calculation, the Doppler factor of the jet decreases from $\delta_D=13.6$ to $\delta_D =10.9$ after the jet passing through the BLR. Thus, the deceleration of the jet will not significantly influence the leptonic emission from dissipation zones outside the BLR.

\section{Conclusion}
In this work, we proposed a hadronuclear origin of the high-energy event from the BL object TXS~0506+056. The multiwavelength flare coincident with the neutrino event can also be explained under the same framework by invoking a second radiation zone outside the BLR.
Our model predicts one (anti)muon event detected by the IceCube EHE alerts per 100\,days and per 3.8\,yrs for a proton injection spectral index of $s=1.6$ and $s=2.0$, respectively, while only a moderate sub-Eddington jet power is required. We suggested that the event IC-170922A is not a lucky detection once there is a dissipation process taking place inside the BLR. The neutrino emission is not necessarily correlated with the multiwavelength flare but it may be accompanied by a spectrum hardening above a few GeV, which is consistent with the Fermi-LAT observation on TXS~0506+056  within two weeks of the neutrino detection, and it may be used as a test for our model in the future. 
%Our model predicts a high TeV-PeV gamma-ray flux at source, which will be reprocessed into diffuse GeV gamma rays through the EM cascades on the cosmic microwave background (CMB) and extragalactic background light (EBL).
The potential of our model to explain TeV emission of other BL Lac objects will be studied and the results can be used to forecast their contributions to the diffuse gamma-ray background and the diffuse high-energy neutrino background. %Lastly, we caveat that since flat spectrum radio quasars (FSRQs), which are another kind of blazars, usually show strong emissions from SMBH accrection disks and from BLRs, the photomeson process may be more important than the $pp$ collision in FSRQs even if there are dense clouds. 
%The neutrino and gamma-ray radiation above 10\,GeV is ascribed to the interactions between the the jet launched from the SMBH and the dense clouds in the vicinity of the SMBH. The dense clouds not only provide a target for the $pp$ collision but also absorb the radiation from UV band to X-ray band in the region. Thus, a second radiation zone, i.e., a radiating blob, is  below 10\,GeV is attributed to a radiating blob at a larger distance to the SMBH. 
%while another shock expands into the cloud with a speed roughly equal to the sound speed in the shocked wind $v_{\rm sc}\sim c\sqrt{\Gamma_j\rho_j/m_Hn_H}$ \citep{Bosch-Ramon12}, where $m_p$ is the proton mass, $c$ is the speed of light and $\rho_j\simeq L_j/\pi R_j^2\Gamma_jv_jc^2$ is the jet mass density with $R_j$ being the jet radius. Protons can 

\acknowledgements We thank Kohta Murase and Markus Ackermann for valuable comments. This work is partially supported by 973 program grant 2014CB845800 and the NSFC grant 11625312 and 11851304.

\clearpage

\appendix

\begin{widetext}

\section{Cascade emission initiated by $pp$ collisions}
The high-energy photons and electrons/positrons (hereafter we do not distinguish positrons from electrons) produced in $pp$ collisions will initiate EM cascades in the BLR via the synchrotron radiation, the inverse Compton (IC) scattering and $\gamma\gamma$ annihilation. As we can see in Fig.~\ref{fig:blr_timescales}, the timescales of these cooling processes are shorter than the dynamical timescale, so we follow the treatment in \citet{Boettcher13, Wang18} for fast-cooling electrons which are assumed to be in quasi-steady state. Assuming a homogeneous spatial distribution of electrons in the BLR, the cascade equation for electrons is given by
\begin{equation}\label{eq:e_traneq}
\frac{\partial N_e'}{\partial t'}+\frac{\partial}{\partial \gamma_e'}\left(\dot{\gamma'_e}N_e'\right)=Q_{e,\pi}'+Q_{e,\gamma\gamma}'-\frac{N_e'}{t_{e,\rm esc}'},
\end{equation}
where 
\begin{equation}
\dot{\gamma_e}'=-\frac{4c\sigma_T}{3 m_ec^2}\left(\frac{B_j'^2}{8\pi}+\Gamma_j^2c_{\rm BLR}aT_{\rm BLR}^4\kappa_{\rm KN}(\gamma_e')\right)\gamma_e'^2
\end{equation}
is the energy loss rate of electrons due to the synchrotron radiation in the magnetic field of the jet and due to the IC radiation in the relativistic boosted photon field of the BLR. In the above equation, $\sigma_T$ is the Thomson cross section, $\kappa_{\rm KN}$ is a numerical factor considering modification of the Klein-Nishina effect to the energy loss rate. We here neglect the electron cooling due to bremsstrahlung radiation, since the cooling time of this process\citep{Ginzburg64} $t_{\rm brem}=6.3\times 10^{5}(n_{\rm H}'/2\times 10^9\,\rm cm^{-3})^{-1}\,$s is much longer than the synchrotron or IC cooling timescale. $t_{e,\rm esc}'$ is the escape timescale of electrons from the BLR (or the residence timescale in the BLR), which is assume to be the dynamic timescale $t_{\rm dyn}'$. On the right-hand side of the equation, $Q_{e,\pi}'=\dot{N}_e'$ represents the injection of electrons from the $pp$ collision via pion decay and $Q_{e,\gamma\gamma}'$ is the injection rate of electrons from $\gamma\gamma$ annihilation of gamma-ray photons, including the annihilation of the high-energy photons from the neutral pion decay produced in the $pp$ collision, and the high-energy photons produced by the synchrotron and the IC radiation, i.e.,
\begin{equation}\label{eq:ggab}
\begin{split}
Q_{e,\gamma\gamma}(\gamma_e')' &= f_{\rm abs}(E_{\gamma,1}')\left(\dot{n}_{E_{\gamma,1}'}^0 + \dot{n}_{E_{\gamma,1}'}^{\rm sy} + \dot{n}_{E_{\gamma,1}'}^{\rm IC}\right)\\
& + f_{\rm abs}(E_{\gamma,2}')\left(\dot{n}_{E_{\gamma,2}'}^0 + \dot{n}_{E_{\gamma,2}'}^{\rm sy} + \dot{n}_{E_{\gamma,2}'}^{\rm IC}\right),
\end{split}
\end{equation}
with
\begin{equation}\label{eq:esc_frac}
f_{\rm abs}(E_\gamma') = 1 - \frac{1-e^{-\tau_{\gamma\gamma}(E_\gamma')}}{\tau _{\gamma\gamma}(E_\gamma')}
\end{equation}
being the absorbed fraction of photons. $\tau_{\gamma\gamma}$ is the optical depth of the high-energy photon of energy $E_\gamma'$ due to $\gamma\gamma$ annihilation. Since the optical depth is a Lorentz invariant, we calculate it in the source frame by 
\begin{equation}
\tau_{\gamma\gamma}(E_\gamma')=\frac{2R_{\rm BLR}}{E_\gamma}\int_1^{\infty}s\sigma_{\gamma\gamma}(s)\int_{sm_e^2c^4/2E_\gamma}^{\infty}\frac{n_{\rm ph}(\varepsilon)}{\varepsilon^2}d\varepsilon
\end{equation}
where $E_\gamma=\Gamma_jE_\gamma'$, $\sqrt{s}$ is the center-of-momentum Lorentz factor of the produced pair, $\varepsilon$ is the photon energy of the BLR and $\sigma_{\gamma\gamma}$ is the total cross section for the $\gamma\gamma$ annihilation given by \citep{Aharonian83_epp}.

Two electrons are produced in each $\gamma\gamma$ annihilation, taking a fraction of $f_\gamma$ and $1-f_\gamma$ of the energy of the incident gamma-ray photon, respectively. Therefore, to the produce an electron with energy $\gamma_e'$, the photons need to have the energy of either $E_{\gamma,1}'= \gamma_e'/f_\gamma$, or $E_{\gamma,2}' = \gamma_e'/(1 - f_\gamma)$. That is the reason why Eq.~(\ref{eq:ggab}) contains two parts. According to \citet{Boettcher13}, taking $f_\gamma=0.9$ can lead to a cascade spectrum in a good agreement with the numerical Monte Carlo simulations. 

In the quasi-steady state, we have $\frac{\partial N_e'}{\partial t}=0$ and the solution to Eq.~\ref{eq:e_traneq} is given by
\begin{equation}\label{eq:sol_e}
N'_e(\gamma_e') = -\frac{1}{\dot{\gamma}'_e}\int_{\gamma_e'}^\infty  d \tilde{\gamma}_e'\left[Q_e(\tilde{\gamma}_e') + \dot{N}_{e,\gamma\gamma}' (\tilde{\gamma}_e')-\frac{N'_e(\tilde{\gamma}_e')}{t_{e,\rm esc}'}\right],
\end{equation}
Since the electron spectrum ${N'}_e(\gamma_e')$ appears at both sides of the Eq.~(\ref{eq:sol_e}), the electron spectrum is calculated progressively, namely, starting from the highest electron energies and then using the solution of $N_e'(\gamma_e')$ for large $\gamma_e'$ as one progress toward the lower values of $\gamma_e'$, to obtain the final electron spectrum in the quasi-steady state. The obtained electron energy spectrum in the jet comoving frame is shown in Fig.~\ref{fig:cas_e_spec}. Then, we use the obtained $N_e'$ to get the synchrotron and IC radiation of cascaded electrons in the quasi-steady state. In Fig.~\ref{fig:decompose}, we decompose the hadronic emission in the BLR into difference components.

\begin{figure}[htbp]
\centering
\includegraphics[width=0.9\columnwidth]{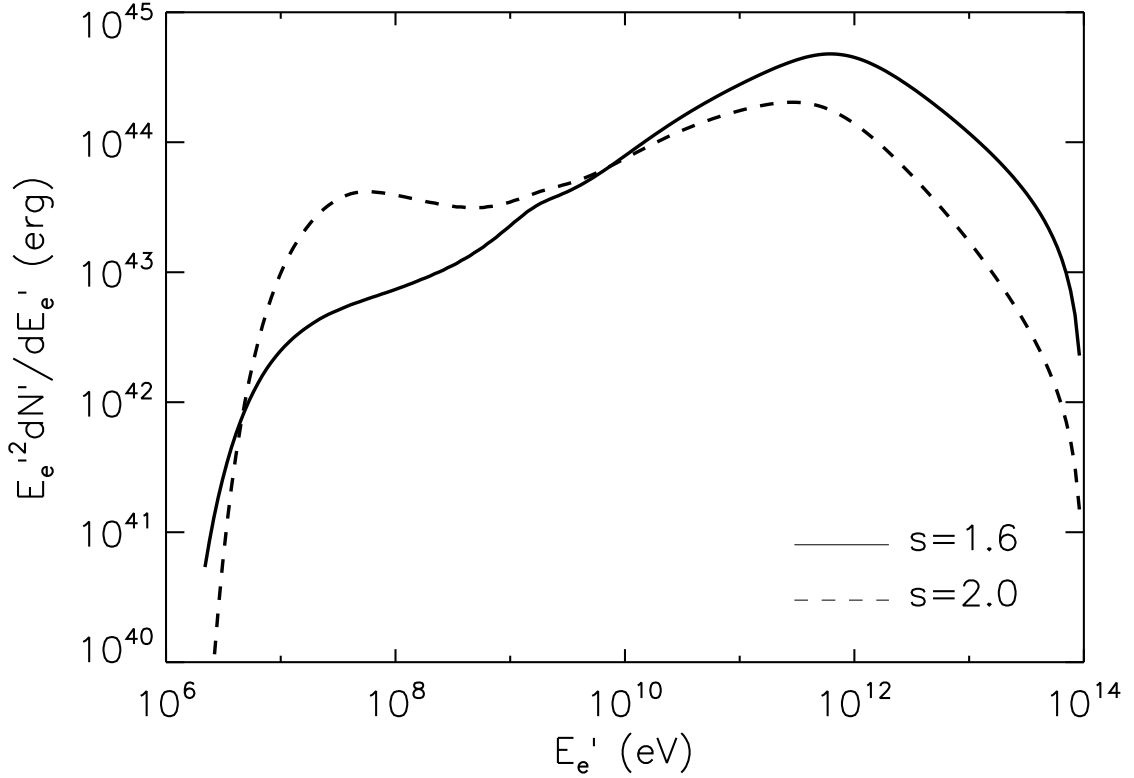}
\caption{Steady-state electron energy spectrum in the cascade. The solid and the dashed curves are for $s=1.6$ and $s=2.0$ respectively.}\label{fig:cas_e_spec}
\end{figure}

\begin{figure}[htbp]
\centering
\includegraphics[width=0.9\columnwidth]{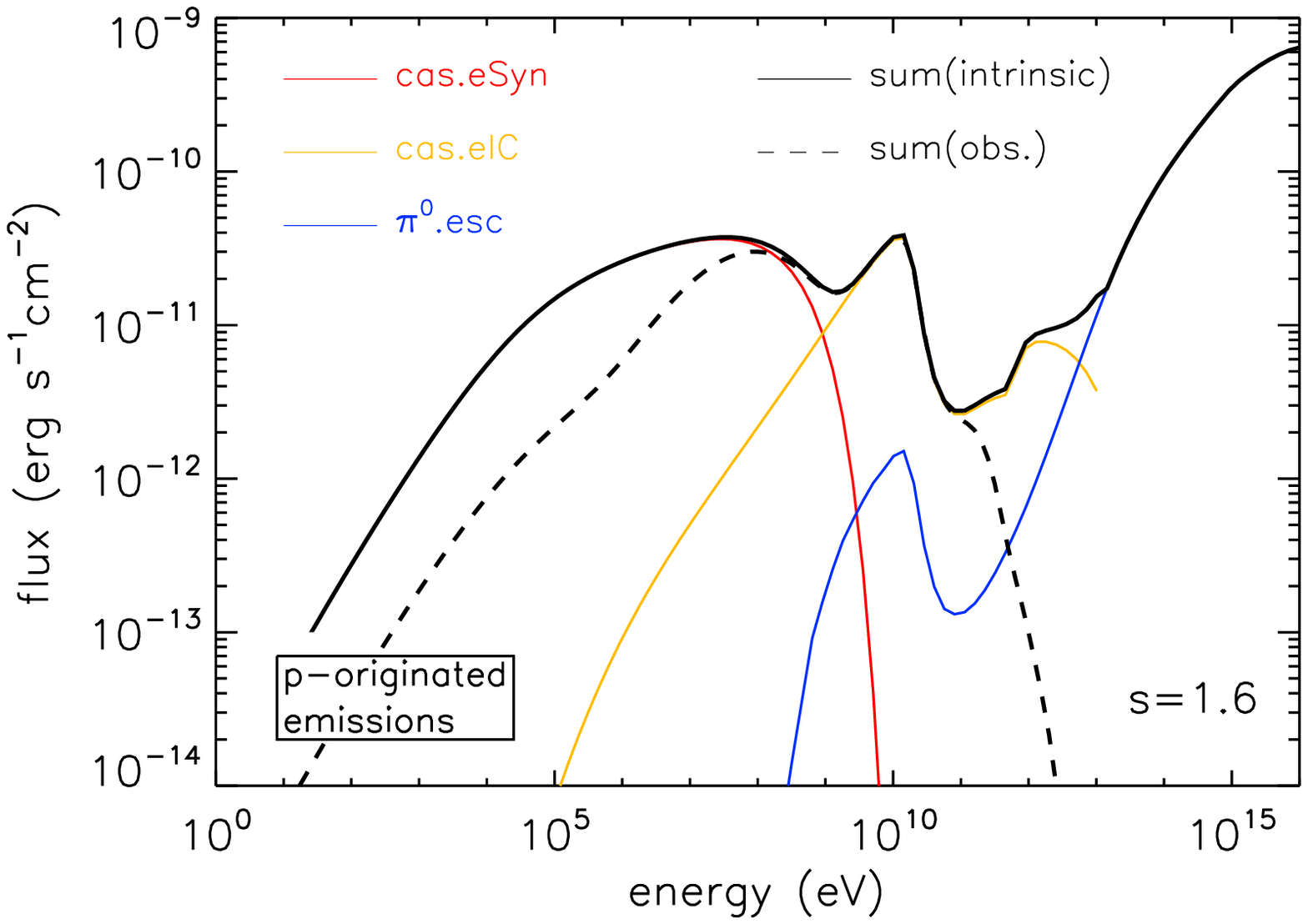}
\includegraphics[width=0.9\columnwidth]{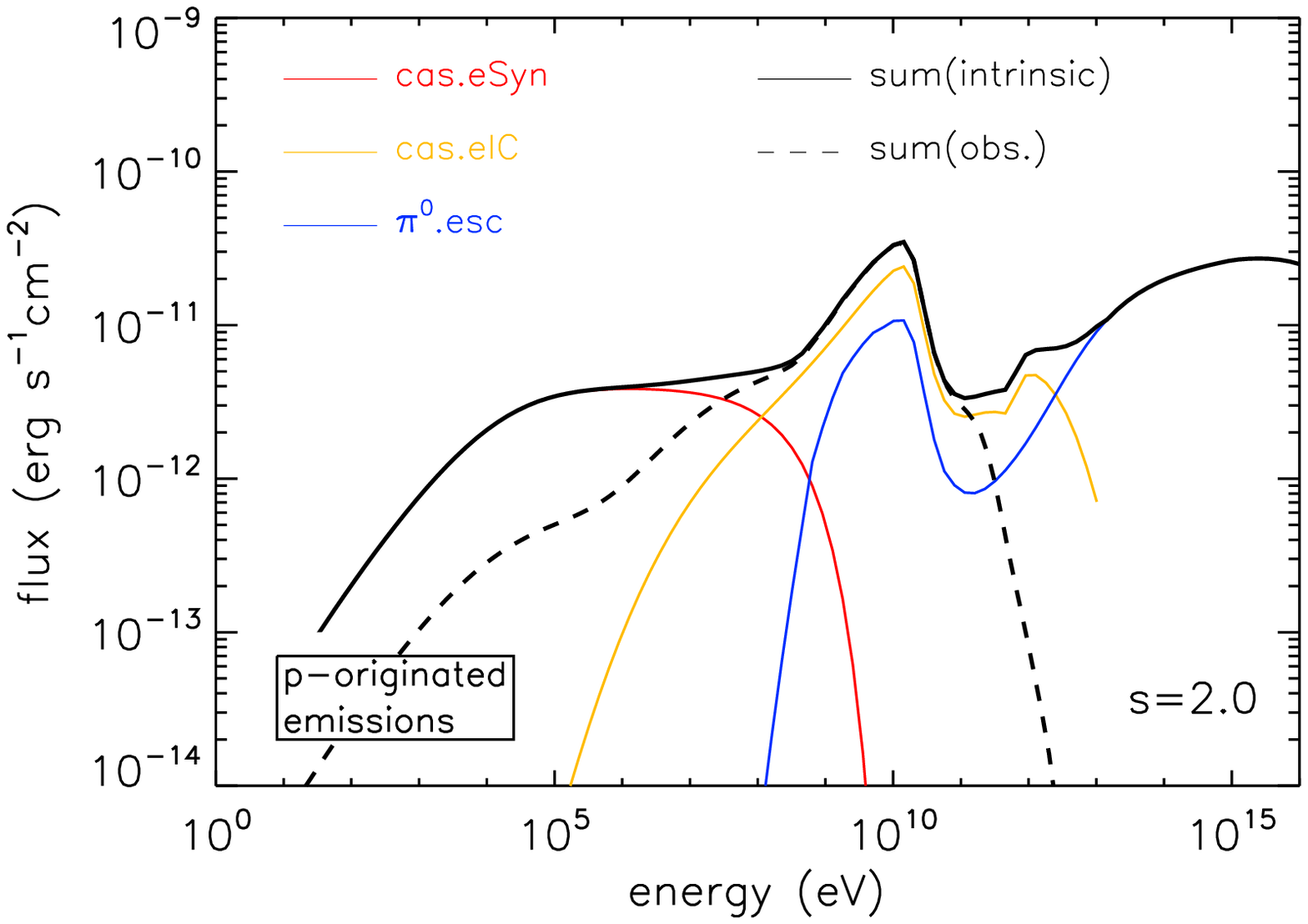}
\caption{Fluxes of various hadronic-originated emissions in the BLR. The red and orange solid curves represent, respectively, the synchrotron and inverse Compton radiation of electrons generated in the cascade. The blue curve represents escaping pionic gamma rays (i.e., gamma rays that are not annihilated and hence do not attend the cascade process). The black solid curve are the summon of the above three components and the black dashed curves are the flux after absorption through photoionisation and due to EBL absorption. The upper panel is for $s=1.6$ and the lower panel is for $s=2.0$.}\label{fig:decompose}
\end{figure}

\subsection{Influence of an infrared photon field from dusty torus}
{Now let us study the effect of an additional infrared photon field supplied by the possibly existed dusty torus. The dusty torus generally locate at an extended region of $0.1-10$pc. Similar to the BLR, the torus absorbs part of the AGN emission and reprocesses it into infrared emissions, which consist of multiple grey body components of temperature ranging from $\sim 50\,$K to $1000\,$K. High-energy gamma rays that escape the BLR may interact with the infrared photon field of the dusty torus, generate electron pairs and re-emit at lower energy. Let us consider that the dust of temperature $T_{\rm DT}$ emit at a luminosity of $L_{\rm DT}$, extending a spatial scale of $R_{\rm DT}$. They supply a photon field of number density 
\begin{equation}
n_{\rm ph, DT}\sim \frac{L_{\rm IR}}{3kT_{\rm DT}\pi R_{\rm DT}^2c}\simeq 8\times 10^5(L_{\rm IR}/10^{41}{\rm erg~s^{-1}})(R_{\rm DT}/1\,{\rm pc})^{-2}(T_{\rm DT}/300\rm \,K)^{-1}\,cm^{-3}
\end{equation}
within a scale of $R_{\rm DT}$ around the SMBH. For hot dust of temperature $1000\,K$ extending a spatial scale of $R_{\rm DT}=0.1$\,pc, we obtain a photon number density of $n_{\rm ph, DT}=3\times 10^7\,\rm cm^{-3}$ with $L_{\rm DT}=10^{41}\,$erg/s which is comparable to the BLR luminosity. Such an infrared photon field typically absorbs $\sim 4\,$TeV gamma rays. The optical depth of $\gamma\gamma$ annihilation can be estimated by $\tau_{\gamma\gamma, \rm DT}\simeq n_{\rm ph, DT}\sigma_{\rm \gamma\gamma}R_{\rm DT}\simeq 0.8$ for gamma-ray photons of energy 4\,TeV typically. The photon density from hot dust will drop quickly and become anisotropic at the region beyond 0.1\,pc and hence do not further contribute the optical depth. Similarly, we can obtain the optical depth by warm dust of 300\,K at a scale of 1\,pc by $\tau_{\gamma\gamma,\rm  DT}\simeq 0.2$ for $\sim$10\,TeV gamma rays and by cold dust of 50\,K at a scale of 10\,pc by $\tau_{\gamma\gamma,\rm  DT}\simeq 0.1$ for $\sim$80\,TeV gamma rays, if we assume the luminosity of each of these emitters is $L_{\rm DT}=10^{41}\,$erg/s. Therefore, only a small fraction of the energy of escaping gamma rays will go into lower energy emission and do not add to the jet emission. 

Assuming the infrared photon field is composed of grey body emissions of the dusty torus of three temperatures at different spatial scales, we employ Eqs.(\ref{eq:e_traneq})-(\ref{eq:sol_e}) to deal with the cascade emission in the infrared photon field with a few modifications: (i) for the electron injection, the term $Q_{e,\pi}'$ will not show up in Eq.~(\ref{eq:e_traneq}) since there is no target for $pp$ collision beyond the BLR; (ii) Eq.~(\ref{eq:ggab}) now reads
\begin{equation}
\begin{split}
Q_{e,\gamma\gamma}^{\rm DT}(\gamma_e')' &= f_{\rm abs}^{\rm DT}(E_{\gamma,1}')\left( \dot{n}_{E_{\gamma,1}'}^{\rm sy} + \dot{n}_{E_{\gamma,1}'}^{\rm IC}\right)+g_{\rm abs}^{\rm DT}\dot{n}_{E_{\gamma,1}'}^{\rm BLR}\\
& + f_{\rm abs}^{\rm DT}(E_{\gamma,2}')\left(\dot{n}_{E_{\gamma,2}'}^{\rm sy} + \dot{n}_{E_{\gamma,2}'}^{\rm IC}\right)+g_{\rm abs}^{\rm DT}\dot{n}_{E_{\gamma,2}'}^{\rm BLR},
\end{split}
\end{equation}
where $\dot{n}_{E_{\gamma, 1/2}'}^{\rm BLR}$ is the photon emission rate of the BLR obtained above. $f_{\rm abs}^{\rm DT}$ holds the same form of Eq.~\ref{eq:esc_frac}, while $g_{\rm abs}^{\rm DT}=1-e^{-\tau_{\gamma\gamma,DT}}$ because photons injected from the BLR will penetrate the whole infrared photon field; (iii) for cascade emission in the infrared photon field, there is no opacity from ionised electrons since the region where the cascade develops is far beyond the BLR. We compare photon fluxes obtained with and without considering the emission of dusty torus in Fig.~\ref{fig:compare_IR}. As can be seen, the predicted flux almost does not change after introducing the infrared emission from the dusty torus. The magnetic field in this extended region of of $R=0.1-10\,$pc is supposed to be much weaker than that in the BLR since this region is far beyond the dissipation region. We the magnetic field density decrease as $R^{-2}$ (i.e., magnetic luminosity conserves) in the calculation. At a larger spatial scale of $\sim 100\,$pc, the escaping gamma rays can be absorbed by the reprocessed emissions of dust in the starburst region of the host galaxy. We assume the generated electrons will be isotropised and their emission is negligible compared to the jet emission. }

\begin{figure}[htbp]
\centering
\includegraphics[width=0.9\columnwidth]{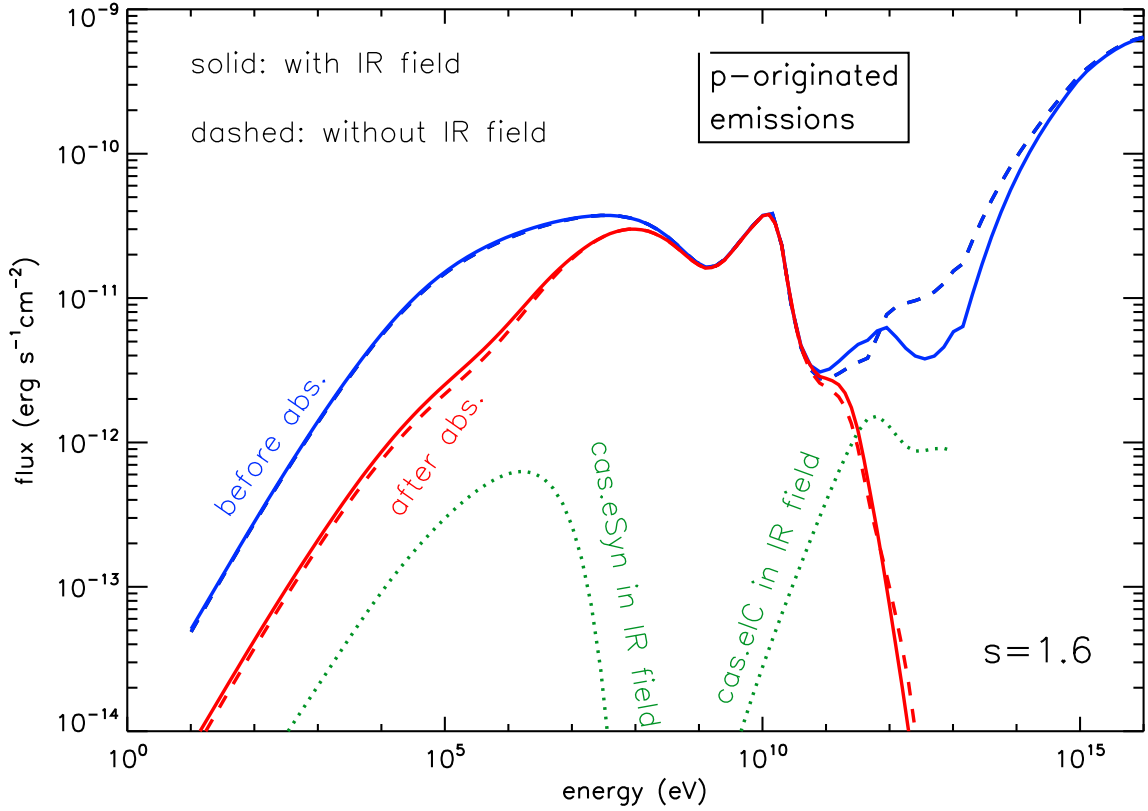}
\caption{Comparison of hadronic-originated fluxes between the case with (solid curves) considering the infrared photon field emitted by the dusty torus and the case without (dashed curves) considering it. The red curves represent the fluxes after the attenuation due to EBL (at high-energy end) and due to Compton scattering of the electrons in the BLR (at low-energy end), while the blue curves represent the flux before the attenuation. The green dotted curve shows the synchrotron radiation and IC radiation of electrons generated in the infrared photon field.}\label{fig:compare_IR}
\end{figure}

\end{widetext}

\bibliographystyle{apsrev4-1}
\bibliography{ms_prd.bib}

\begin{thebibliography}{64}%
\makeatletter
\providecommand \@ifxundefined [1]{%
 \@ifx{#1\undefined}
}%
\providecommand \@ifnum [1]{%
 \ifnum #1\expandafter \@firstoftwo
 \else \expandafter \@secondoftwo
 \fi
}%
\providecommand \@ifx [1]{%
 \ifx #1\expandafter \@firstoftwo
 \else \expandafter \@secondoftwo
 \fi
}%
\providecommand \natexlab [1]{#1}%
\providecommand \enquote  [1]{``#1''}%
\providecommand \bibnamefont  [1]{#1}%
\providecommand \bibfnamefont [1]{#1}%
\providecommand \citenamefont [1]{#1}%
\providecommand \href@noop [0]{\@secondoftwo}%
\providecommand \href [0]{\begingroup \@sanitize@url \@href}%
\providecommand \@href[1]{\@@startlink{#1}\@@href}%
\providecommand \@@href[1]{\endgroup#1\@@endlink}%
\providecommand \@sanitize@url [0]{\catcode `\\12\catcode `\$12\catcode
  `\&12\catcode `\#12\catcode `\^12\catcode `\_12\catcode `\%12\relax}%
\providecommand \@@startlink[1]{}%
\providecommand \@@endlink[0]{}%
\providecommand \url  [0]{\begingroup\@sanitize@url \@url }%
\providecommand \@url [1]{\endgroup\@href {#1}{\urlprefix }}%
\providecommand \urlprefix  [0]{URL }%
\providecommand \Eprint [0]{\href }%
\providecommand \doibase [0]{http://dx.doi.org/}%
\providecommand \selectlanguage [0]{\@gobble}%
\providecommand \bibinfo  [0]{\@secondoftwo}%
\providecommand \bibfield  [0]{\@secondoftwo}%
\providecommand \translation [1]{[#1]}%
\providecommand \BibitemOpen [0]{}%
\providecommand \bibitemStop [0]{}%
\providecommand \bibitemNoStop [0]{.\EOS\space}%
\providecommand \EOS [0]{\spacefactor3000\relax}%
\providecommand \BibitemShut  [1]{\csname bibitem#1\endcsname}%
\let\auto@bib@innerbib\@empty
%</preamble>
\bibitem [{\citenamefont {{IceCube Collaboration}}(2013)}]{IC13_sci}%
  \BibitemOpen
  \bibfield  {author} {\bibinfo {author} {\bibnamefont {{IceCube
  Collaboration}}},\ }\href {\doibase 10.1126/science.1242856} {\bibfield
  {journal} {\bibinfo  {journal} {Science}\ }\textbf {\bibinfo {volume}
  {342}},\ \bibinfo {eid} {1242856} (\bibinfo {year} {2013})},\ \Eprint
  {http://arxiv.org/abs/1311.5238} {arXiv:1311.5238 [astro-ph.HE]} \BibitemShut
  {NoStop}%
\bibitem [{\citenamefont {{Liu}}\ \emph {et~al.}(2014)\citenamefont {{Liu}},
  \citenamefont {{Wang}}, \citenamefont {{Inoue}}, \citenamefont {{Crocker}},\
  and\ \citenamefont {{Aharonian}}}]{Liu14}%
  \BibitemOpen
  \bibfield  {author} {\bibinfo {author} {\bibfnamefont {R.-Y.}\ \bibnamefont
  {{Liu}}}, \bibinfo {author} {\bibfnamefont {X.-Y.}\ \bibnamefont {{Wang}}},
  \bibinfo {author} {\bibfnamefont {S.}~\bibnamefont {{Inoue}}}, \bibinfo
  {author} {\bibfnamefont {R.}~\bibnamefont {{Crocker}}}, \ and\ \bibinfo
  {author} {\bibfnamefont {F.}~\bibnamefont {{Aharonian}}},\ }\href {\doibase
  10.1103/PhysRevD.89.083004} {\bibfield  {journal} {\bibinfo  {journal}
  {\prd}\ }\textbf {\bibinfo {volume} {89}},\ \bibinfo {eid} {083004} (\bibinfo
  {year} {2014})},\ \Eprint {http://arxiv.org/abs/1310.1263} {arXiv:1310.1263
  [astro-ph.HE]} \BibitemShut {NoStop}%
\bibitem [{\citenamefont {{Tamborra}}\ \emph {et~al.}(2014)\citenamefont
  {{Tamborra}}, \citenamefont {{Ando}},\ and\ \citenamefont
  {{Murase}}}]{Tamborra14}%
  \BibitemOpen
  \bibfield  {author} {\bibinfo {author} {\bibfnamefont {I.}~\bibnamefont
  {{Tamborra}}}, \bibinfo {author} {\bibfnamefont {S.}~\bibnamefont {{Ando}}},
  \ and\ \bibinfo {author} {\bibfnamefont {K.}~\bibnamefont {{Murase}}},\
  }\href {\doibase 10.1088/1475-7516/2014/09/043} {\bibfield  {journal}
  {\bibinfo  {journal} {\jcap}\ }\textbf {\bibinfo {volume} {9}},\ \bibinfo
  {eid} {043} (\bibinfo {year} {2014})},\ \Eprint
  {http://arxiv.org/abs/1404.1189} {arXiv:1404.1189 [astro-ph.HE]} \BibitemShut
  {NoStop}%
\bibitem [{\citenamefont {{Chang}}\ \emph {et~al.}(2015)\citenamefont
  {{Chang}}, \citenamefont {{Liu}},\ and\ \citenamefont {{Wang}}}]{Chang15}%
  \BibitemOpen
  \bibfield  {author} {\bibinfo {author} {\bibfnamefont {X.-C.}\ \bibnamefont
  {{Chang}}}, \bibinfo {author} {\bibfnamefont {R.-Y.}\ \bibnamefont {{Liu}}},
  \ and\ \bibinfo {author} {\bibfnamefont {X.-Y.}\ \bibnamefont {{Wang}}},\
  }\href {\doibase 10.1088/0004-637X/805/2/95} {\bibfield  {journal} {\bibinfo
  {journal} {\apj}\ }\textbf {\bibinfo {volume} {805}},\ \bibinfo {eid} {95}
  (\bibinfo {year} {2015})},\ \Eprint {http://arxiv.org/abs/1412.8361}
  {arXiv:1412.8361 [astro-ph.HE]} \BibitemShut {NoStop}%
\bibitem [{\citenamefont {{Wang}}\ and\ \citenamefont {{Liu}}(2016)}]{Wang16}%
  \BibitemOpen
  \bibfield  {author} {\bibinfo {author} {\bibfnamefont {X.-Y.}\ \bibnamefont
  {{Wang}}}\ and\ \bibinfo {author} {\bibfnamefont {R.-Y.}\ \bibnamefont
  {{Liu}}},\ }\href {\doibase 10.1103/PhysRevD.93.083005} {\bibfield  {journal}
  {\bibinfo  {journal} {\prd}\ }\textbf {\bibinfo {volume} {93}},\ \bibinfo
  {eid} {083005} (\bibinfo {year} {2016})},\ \Eprint
  {http://arxiv.org/abs/1512.08596} {arXiv:1512.08596 [astro-ph.HE]}
  \BibitemShut {NoStop}%
\bibitem [{\citenamefont {{Lunardini}}\ and\ \citenamefont
  {{Winter}}(2017)}]{Lunardini17}%
  \BibitemOpen
  \bibfield  {author} {\bibinfo {author} {\bibfnamefont {C.}~\bibnamefont
  {{Lunardini}}}\ and\ \bibinfo {author} {\bibfnamefont {W.}~\bibnamefont
  {{Winter}}},\ }\href {\doibase 10.1103/PhysRevD.95.123001} {\bibfield
  {journal} {\bibinfo  {journal} {\prd}\ }\textbf {\bibinfo {volume} {95}},\
  \bibinfo {eid} {123001} (\bibinfo {year} {2017})},\ \Eprint
  {http://arxiv.org/abs/1612.03160} {arXiv:1612.03160 [astro-ph.HE]}
  \BibitemShut {NoStop}%
\bibitem [{\citenamefont {{Senno}}\ \emph {et~al.}(2017)\citenamefont
  {{Senno}}, \citenamefont {{Murase}},\ and\ \citenamefont
  {{M{\'e}sz{\'a}ros}}}]{Senno17}%
  \BibitemOpen
  \bibfield  {author} {\bibinfo {author} {\bibfnamefont {N.}~\bibnamefont
  {{Senno}}}, \bibinfo {author} {\bibfnamefont {K.}~\bibnamefont {{Murase}}}, \
  and\ \bibinfo {author} {\bibfnamefont {P.}~\bibnamefont
  {{M{\'e}sz{\'a}ros}}},\ }\href {\doibase 10.3847/1538-4357/aa6344} {\bibfield
   {journal} {\bibinfo  {journal} {\apj}\ }\textbf {\bibinfo {volume} {838}},\
  \bibinfo {eid} {3} (\bibinfo {year} {2017})},\ \Eprint
  {http://arxiv.org/abs/1612.00918} {arXiv:1612.00918 [astro-ph.HE]}
  \BibitemShut {NoStop}%
\bibitem [{\citenamefont {{Stecker}}(2013)}]{Stecker13}%
  \BibitemOpen
  \bibfield  {author} {\bibinfo {author} {\bibfnamefont {F.~W.}\ \bibnamefont
  {{Stecker}}},\ }\href {\doibase 10.1103/PhysRevD.88.047301} {\bibfield
  {journal} {\bibinfo  {journal} {\prd}\ }\textbf {\bibinfo {volume} {88}},\
  \bibinfo {eid} {047301} (\bibinfo {year} {2013})},\ \Eprint
  {http://arxiv.org/abs/1305.7404} {arXiv:1305.7404 [astro-ph.HE]} \BibitemShut
  {NoStop}%
\bibitem [{\citenamefont {{Murase}}\ \emph {et~al.}(2014)\citenamefont
  {{Murase}}, \citenamefont {{Inoue}},\ and\ \citenamefont
  {{Dermer}}}]{Murase14}%
  \BibitemOpen
  \bibfield  {author} {\bibinfo {author} {\bibfnamefont {K.}~\bibnamefont
  {{Murase}}}, \bibinfo {author} {\bibfnamefont {Y.}~\bibnamefont {{Inoue}}}, \
  and\ \bibinfo {author} {\bibfnamefont {C.~D.}\ \bibnamefont {{Dermer}}},\
  }\href {\doibase 10.1103/PhysRevD.90.023007} {\bibfield  {journal} {\bibinfo
  {journal} {\prd}\ }\textbf {\bibinfo {volume} {90}},\ \bibinfo {eid} {023007}
  (\bibinfo {year} {2014})},\ \Eprint {http://arxiv.org/abs/1403.4089}
  {arXiv:1403.4089 [astro-ph.HE]} \BibitemShut {NoStop}%
\bibitem [{\citenamefont {{Tavecchio}}\ and\ \citenamefont
  {{Ghisellini}}(2015)}]{Tavecchio15}%
  \BibitemOpen
  \bibfield  {author} {\bibinfo {author} {\bibfnamefont {F.}~\bibnamefont
  {{Tavecchio}}}\ and\ \bibinfo {author} {\bibfnamefont {G.}~\bibnamefont
  {{Ghisellini}}},\ }\href {\doibase 10.1093/mnras/stv1023} {\bibfield
  {journal} {\bibinfo  {journal} {\mnras}\ }\textbf {\bibinfo {volume} {451}},\
  \bibinfo {pages} {1502} (\bibinfo {year} {2015})},\ \Eprint
  {http://arxiv.org/abs/1411.2783} {arXiv:1411.2783 [astro-ph.HE]} \BibitemShut
  {NoStop}%
\bibitem [{\citenamefont {{Petropoulou}}\ \emph {et~al.}(2015)\citenamefont
  {{Petropoulou}}, \citenamefont {{Dimitrakoudis}}, \citenamefont {{Padovani}},
  \citenamefont {{Mastichiadis}},\ and\ \citenamefont
  {{Resconi}}}]{Petropoulou15}%
  \BibitemOpen
  \bibfield  {author} {\bibinfo {author} {\bibfnamefont {M.}~\bibnamefont
  {{Petropoulou}}}, \bibinfo {author} {\bibfnamefont {S.}~\bibnamefont
  {{Dimitrakoudis}}}, \bibinfo {author} {\bibfnamefont {P.}~\bibnamefont
  {{Padovani}}}, \bibinfo {author} {\bibfnamefont {A.}~\bibnamefont
  {{Mastichiadis}}}, \ and\ \bibinfo {author} {\bibfnamefont {E.}~\bibnamefont
  {{Resconi}}},\ }\href {\doibase 10.1093/mnras/stv179} {\bibfield  {journal}
  {\bibinfo  {journal} {\mnras}\ }\textbf {\bibinfo {volume} {448}},\ \bibinfo
  {pages} {2412} (\bibinfo {year} {2015})},\ \Eprint
  {http://arxiv.org/abs/1501.07115} {arXiv:1501.07115 [astro-ph.HE]}
  \BibitemShut {NoStop}%
\bibitem [{\citenamefont {{Padovani}}\ \emph {et~al.}(2015)\citenamefont
  {{Padovani}}, \citenamefont {{Petropoulou}}, \citenamefont {{Giommi}},\ and\
  \citenamefont {{Resconi}}}]{Padovani15}%
  \BibitemOpen
  \bibfield  {author} {\bibinfo {author} {\bibfnamefont {P.}~\bibnamefont
  {{Padovani}}}, \bibinfo {author} {\bibfnamefont {M.}~\bibnamefont
  {{Petropoulou}}}, \bibinfo {author} {\bibfnamefont {P.}~\bibnamefont
  {{Giommi}}}, \ and\ \bibinfo {author} {\bibfnamefont {E.}~\bibnamefont
  {{Resconi}}},\ }\href {\doibase 10.1093/mnras/stv1467} {\bibfield  {journal}
  {\bibinfo  {journal} {\mnras}\ }\textbf {\bibinfo {volume} {452}},\ \bibinfo
  {pages} {1877} (\bibinfo {year} {2015})},\ \Eprint
  {http://arxiv.org/abs/1506.09135} {arXiv:1506.09135 [astro-ph.HE]}
  \BibitemShut {NoStop}%
\bibitem [{\citenamefont {{Kopper}}\ and\ \citenamefont
  {{Blaufuss}}(2017)}]{IC17_GCN}%
  \BibitemOpen
  \bibfield  {author} {\bibinfo {author} {\bibfnamefont {C.}~\bibnamefont
  {{Kopper}}}\ and\ \bibinfo {author} {\bibfnamefont {E.}~\bibnamefont
  {{Blaufuss}}},\ }\href@noop {} {\bibfield  {journal} {\bibinfo  {journal}
  {GRB Coordinates Network, Circular Service, No.~21916, \#1 (2017)}\ }\textbf
  {\bibinfo {volume} {21916}} (\bibinfo {year} {2017})}\BibitemShut {NoStop}%
\bibitem [{\citenamefont {{The IceCube}}\ \emph {et~al.}(2018)\citenamefont
  {{The IceCube}}, \citenamefont {{\emph{Fermi}-LAT}}, \citenamefont {{MAGIC}},
  \citenamefont {{\emph{AGILE}}}, \citenamefont {{ASAS-SN}}, \citenamefont
  {{HAWC}}, \citenamefont {{H.E.S.S.}}, \citenamefont {{\emph{INTEGRAL}}},
  \citenamefont {{Kanata}}, \citenamefont {{Kiso}}, \citenamefont {{Kapteyn}},
  \citenamefont {{Liverpool telescope}}, \citenamefont {{Subaru}},
  \citenamefont {{\emph{Swift/NuSTAR}}}, \citenamefont {{VERITAS}},\ and\
  \citenamefont {{VLA/17B-403 teams}}}]{IC18_TXS}%
  \BibitemOpen
  \bibfield  {author} {\bibinfo {author} {\bibnamefont {{The IceCube}}},
  \bibinfo {author} {\bibnamefont {{\emph{Fermi}-LAT}}}, \bibinfo {author}
  {\bibnamefont {{MAGIC}}}, \bibinfo {author} {\bibnamefont {{\emph{AGILE}}}},
  \bibinfo {author} {\bibnamefont {{ASAS-SN}}}, \bibinfo {author} {\bibnamefont
  {{HAWC}}}, \bibinfo {author} {\bibnamefont {{H.E.S.S.}}}, \bibinfo {author}
  {\bibnamefont {{\emph{INTEGRAL}}}}, \bibinfo {author} {\bibnamefont
  {{Kanata}}}, \bibinfo {author} {\bibnamefont {{Kiso}}}, \bibinfo {author}
  {\bibnamefont {{Kapteyn}}}, \bibinfo {author} {\bibnamefont {{Liverpool
  telescope}}}, \bibinfo {author} {\bibnamefont {{Subaru}}}, \bibinfo {author}
  {\bibnamefont {{\emph{Swift/NuSTAR}}}}, \bibinfo {author} {\bibnamefont
  {{VERITAS}}}, \ and\ \bibinfo {author} {\bibnamefont {{VLA/17B-403 teams}}},\
  }\href {\doibase 10.1126/science.aat1378} {\bibfield  {journal} {\bibinfo
  {journal} {Science}\ }\textbf {\bibinfo {volume} {361}} (\bibinfo {year}
  {2018}),\ 10.1126/science.aat1378},\ \Eprint
  {http://arxiv.org/abs/http://science.sciencemag.org/content/361/6398/eaat1378.full.pdf}
  {http://science.sciencemag.org/content/361/6398/eaat1378.full.pdf}
  \BibitemShut {NoStop}%
\bibitem [{\citenamefont {{Paiano}}\ \emph {et~al.}(2018)\citenamefont
  {{Paiano}}, \citenamefont {{Falomo}}, \citenamefont {{Treves}},\ and\
  \citenamefont {{Scarpa}}}]{Paiano18}%
  \BibitemOpen
  \bibfield  {author} {\bibinfo {author} {\bibfnamefont {S.}~\bibnamefont
  {{Paiano}}}, \bibinfo {author} {\bibfnamefont {R.}~\bibnamefont {{Falomo}}},
  \bibinfo {author} {\bibfnamefont {A.}~\bibnamefont {{Treves}}}, \ and\
  \bibinfo {author} {\bibfnamefont {R.}~\bibnamefont {{Scarpa}}},\ }\href
  {\doibase 10.3847/2041-8213/aaad5e} {\bibfield  {journal} {\bibinfo
  {journal} {\apjl}\ }\textbf {\bibinfo {volume} {854}},\ \bibinfo {eid} {L32}
  (\bibinfo {year} {2018})},\ \Eprint {http://arxiv.org/abs/1802.01939}
  {arXiv:1802.01939} \BibitemShut {NoStop}%
\bibitem [{\citenamefont {{Tanaka}}\ \emph {et~al.}(2017)\citenamefont
  {{Tanaka}}, \citenamefont {{Buson}},\ and\ \citenamefont
  {{Kocevski}}}]{Fermi17_atel}%
  \BibitemOpen
  \bibfield  {author} {\bibinfo {author} {\bibfnamefont {Y.~T.}\ \bibnamefont
  {{Tanaka}}}, \bibinfo {author} {\bibfnamefont {S.}~\bibnamefont {{Buson}}}, \
  and\ \bibinfo {author} {\bibfnamefont {D.}~\bibnamefont {{Kocevski}}},\
  }\href@noop {} {\bibfield  {journal} {\bibinfo  {journal} {The Astronomer's
  Telegram}\ }\textbf {\bibinfo {volume} {10791}} (\bibinfo {year}
  {2017})}\BibitemShut {NoStop}%
\bibitem [{\citenamefont {{Mirzoyan}}(2017)}]{MAGIC17_atel}%
  \BibitemOpen
  \bibfield  {author} {\bibinfo {author} {\bibfnamefont {R.}~\bibnamefont
  {{Mirzoyan}}},\ }\href@noop {} {\bibfield  {journal} {\bibinfo  {journal}
  {The Astronomer's Telegram}\ }\textbf {\bibinfo {volume} {10817}} (\bibinfo
  {year} {2017})}\BibitemShut {NoStop}%
\bibitem [{\citenamefont {{Fox}}\ \emph {et~al.}(2017)\citenamefont {{Fox}},
  \citenamefont {{DeLaunay}}, \citenamefont {{Keivani}}, \citenamefont
  {{Evans}}, \citenamefont {{Turley}}, \citenamefont {{Kennea}}, \citenamefont
  {{Cowen}}, \citenamefont {{Osborne}}, \citenamefont {{Santander}},\ and\
  \citenamefont {{Marshall}}}]{X17_atel}%
  \BibitemOpen
  \bibfield  {author} {\bibinfo {author} {\bibfnamefont {D.~B.}\ \bibnamefont
  {{Fox}}}, \bibinfo {author} {\bibfnamefont {J.~J.}\ \bibnamefont
  {{DeLaunay}}}, \bibinfo {author} {\bibfnamefont {A.}~\bibnamefont
  {{Keivani}}}, \bibinfo {author} {\bibfnamefont {P.~A.}\ \bibnamefont
  {{Evans}}}, \bibinfo {author} {\bibfnamefont {C.~F.}\ \bibnamefont
  {{Turley}}}, \bibinfo {author} {\bibfnamefont {J.~A.}\ \bibnamefont
  {{Kennea}}}, \bibinfo {author} {\bibfnamefont {D.~F.}\ \bibnamefont
  {{Cowen}}}, \bibinfo {author} {\bibfnamefont {J.~P.}\ \bibnamefont
  {{Osborne}}}, \bibinfo {author} {\bibfnamefont {M.}~\bibnamefont
  {{Santander}}}, \ and\ \bibinfo {author} {\bibfnamefont {F.~E.}\ \bibnamefont
  {{Marshall}}},\ }\href@noop {} {\bibfield  {journal} {\bibinfo  {journal}
  {The Astronomer's Telegram}\ }\textbf {\bibinfo {volume} {10845}} (\bibinfo
  {year} {2017})}\BibitemShut {NoStop}%
\bibitem [{\citenamefont {{Franckowiak}}\ \emph {et~al.}(2017)\citenamefont
  {{Franckowiak}}, \citenamefont {{Stanek}}, \citenamefont {{Kochanek}},
  \citenamefont {{Thompson}}, \citenamefont {{Holoien}}, \citenamefont
  {{Shappee}}, \citenamefont {{Prieto}},\ and\ \citenamefont
  {{Dong}}}]{ASASSN17_atel}%
  \BibitemOpen
  \bibfield  {author} {\bibinfo {author} {\bibfnamefont {A.}~\bibnamefont
  {{Franckowiak}}}, \bibinfo {author} {\bibfnamefont {K.~Z.}\ \bibnamefont
  {{Stanek}}}, \bibinfo {author} {\bibfnamefont {C.~S.}\ \bibnamefont
  {{Kochanek}}}, \bibinfo {author} {\bibfnamefont {T.~A.}\ \bibnamefont
  {{Thompson}}}, \bibinfo {author} {\bibfnamefont {T.~W.-S.}\ \bibnamefont
  {{Holoien}}}, \bibinfo {author} {\bibfnamefont {B.~J.}\ \bibnamefont
  {{Shappee}}}, \bibinfo {author} {\bibfnamefont {J.~L.}\ \bibnamefont
  {{Prieto}}}, \ and\ \bibinfo {author} {\bibfnamefont {S.}~\bibnamefont
  {{Dong}}},\ }\href@noop {} {\bibfield  {journal} {\bibinfo  {journal} {The
  Astronomer's Telegram}\ }\textbf {\bibinfo {volume} {10794}} (\bibinfo {year}
  {2017})}\BibitemShut {NoStop}%
\bibitem [{\citenamefont {{Morokuma}}\ \emph {et~al.}(2017)\citenamefont
  {{Morokuma}}, \citenamefont {{Tanaka}}, \citenamefont {{Ohta}}, \citenamefont
  {{Matsuoka}}, \citenamefont {{Yamashita}},\ and\ \citenamefont
  {{Kato}}}]{Subaru17_atel}%
  \BibitemOpen
  \bibfield  {author} {\bibinfo {author} {\bibfnamefont {T.}~\bibnamefont
  {{Morokuma}}}, \bibinfo {author} {\bibfnamefont {Y.~T.}\ \bibnamefont
  {{Tanaka}}}, \bibinfo {author} {\bibfnamefont {K.}~\bibnamefont {{Ohta}}},
  \bibinfo {author} {\bibfnamefont {Y.}~\bibnamefont {{Matsuoka}}}, \bibinfo
  {author} {\bibfnamefont {T.}~\bibnamefont {{Yamashita}}}, \ and\ \bibinfo
  {author} {\bibfnamefont {N.}~\bibnamefont {{Kato}}},\ }\href@noop {}
  {\bibfield  {journal} {\bibinfo  {journal} {The Astronomer's Telegram}\
  }\textbf {\bibinfo {volume} {10890}} (\bibinfo {year} {2017})}\BibitemShut
  {NoStop}%
\bibitem [{\citenamefont {{Yamanaka}}\ \emph {et~al.}(2017)\citenamefont
  {{Yamanaka}}, \citenamefont {{Tanaka}}, \citenamefont {{Mori}}, \citenamefont
  {{Kawabata}}, \citenamefont {{Utsumi}}, \citenamefont {{Nakaoka}},
  \citenamefont {{Kawabata}},\ and\ \citenamefont
  {{Nagashima}}}]{Kanata17_atel}%
  \BibitemOpen
  \bibfield  {author} {\bibinfo {author} {\bibfnamefont {M.}~\bibnamefont
  {{Yamanaka}}}, \bibinfo {author} {\bibfnamefont {Y.~T.}\ \bibnamefont
  {{Tanaka}}}, \bibinfo {author} {\bibfnamefont {H.}~\bibnamefont {{Mori}}},
  \bibinfo {author} {\bibfnamefont {K.~S.}\ \bibnamefont {{Kawabata}}},
  \bibinfo {author} {\bibfnamefont {Y.}~\bibnamefont {{Utsumi}}}, \bibinfo
  {author} {\bibfnamefont {T.}~\bibnamefont {{Nakaoka}}}, \bibinfo {author}
  {\bibfnamefont {M.}~\bibnamefont {{Kawabata}}}, \ and\ \bibinfo {author}
  {\bibfnamefont {H.}~\bibnamefont {{Nagashima}}},\ }\href@noop {} {\bibfield
  {journal} {\bibinfo  {journal} {The Astronomer's Telegram}\ }\textbf
  {\bibinfo {volume} {10844}} (\bibinfo {year} {2017})}\BibitemShut {NoStop}%
\bibitem [{\citenamefont {{Coleiro}}\ and\ \citenamefont
  {{Chaty}}(2017)}]{VLT17_atel}%
  \BibitemOpen
  \bibfield  {author} {\bibinfo {author} {\bibfnamefont {A.}~\bibnamefont
  {{Coleiro}}}\ and\ \bibinfo {author} {\bibfnamefont {S.}~\bibnamefont
  {{Chaty}}},\ }\href@noop {} {\bibfield  {journal} {\bibinfo  {journal} {The
  Astronomer's Telegram}\ }\textbf {\bibinfo {volume} {10840}} (\bibinfo {year}
  {2017})}\BibitemShut {NoStop}%
\bibitem [{\citenamefont {{Tetarenko}}\ \emph {et~al.}(2017)\citenamefont
  {{Tetarenko}}, \citenamefont {{Sivakoff}}, \citenamefont {{Kimball}},\ and\
  \citenamefont {{Miller-Jones}}}]{VLA17_atel}%
  \BibitemOpen
  \bibfield  {author} {\bibinfo {author} {\bibfnamefont {A.~J.}\ \bibnamefont
  {{Tetarenko}}}, \bibinfo {author} {\bibfnamefont {G.~R.}\ \bibnamefont
  {{Sivakoff}}}, \bibinfo {author} {\bibfnamefont {A.~E.}\ \bibnamefont
  {{Kimball}}}, \ and\ \bibinfo {author} {\bibfnamefont {J.~C.~A.}\
  \bibnamefont {{Miller-Jones}}},\ }\href@noop {} {\bibfield  {journal}
  {\bibinfo  {journal} {The Astronomer's Telegram}\ }\textbf {\bibinfo {volume}
  {10861}} (\bibinfo {year} {2017})}\BibitemShut {NoStop}%
\bibitem [{\citenamefont {{Donea}}\ and\ \citenamefont
  {{Protheroe}}(2003)}]{Donea03}%
  \BibitemOpen
  \bibfield  {author} {\bibinfo {author} {\bibfnamefont {A.-C.}\ \bibnamefont
  {{Donea}}}\ and\ \bibinfo {author} {\bibfnamefont {R.~J.}\ \bibnamefont
  {{Protheroe}}},\ }\href {\doibase 10.1016/S0927-6505(02)00155-X} {\bibfield
  {journal} {\bibinfo  {journal} {Astroparticle Physics}\ }\textbf {\bibinfo
  {volume} {18}},\ \bibinfo {pages} {377} (\bibinfo {year} {2003})},\ \Eprint
  {http://arxiv.org/abs/astro-ph/0202068} {astro-ph/0202068} \BibitemShut
  {NoStop}%
\bibitem [{\citenamefont {{Giommi}}\ \emph {et~al.}(2013)\citenamefont
  {{Giommi}}, \citenamefont {{Padovani}},\ and\ \citenamefont
  {{Polenta}}}]{Giommi13}%
  \BibitemOpen
  \bibfield  {author} {\bibinfo {author} {\bibfnamefont {P.}~\bibnamefont
  {{Giommi}}}, \bibinfo {author} {\bibfnamefont {P.}~\bibnamefont
  {{Padovani}}}, \ and\ \bibinfo {author} {\bibfnamefont {G.}~\bibnamefont
  {{Polenta}}},\ }\href {\doibase 10.1093/mnras/stt305} {\bibfield  {journal}
  {\bibinfo  {journal} {\mnras}\ }\textbf {\bibinfo {volume} {431}},\ \bibinfo
  {pages} {1914} (\bibinfo {year} {2013})},\ \Eprint
  {http://arxiv.org/abs/1302.4331} {arXiv:1302.4331 [astro-ph.HE]} \BibitemShut
  {NoStop}%
\bibitem [{Note1()}]{Note1}%
  \BibitemOpen
  \bibinfo {note} {During the review period of this paper, we noticed a new
  posted paper on arXiv suggesting that TXS~0506+056 is a ``masquerading'' BL
  Lac and an intrinsically flat-spectrum radio quasar with hidden BLR, see
  P.~Padovani, F.~Oikonomou, M.~Petropoulou, P.~Giommi, and E.~Resconi, Mon.
  Not. R. Astron. Soc. {\protect \bf 484}, 104 (2019).}\BibitemShut {Stop}%
\bibitem [{\citenamefont {{Vermeulen}}\ \emph {et~al.}(1995)\citenamefont
  {{Vermeulen}}, \citenamefont {{Ogle}}, \citenamefont {{Tran}}, \citenamefont
  {{Browne}}, \citenamefont {{Cohen}}, \citenamefont {{Readhead}},
  \citenamefont {{Taylor}},\ and\ \citenamefont {{Goodrich}}}]{Vermeulen95}%
  \BibitemOpen
  \bibfield  {author} {\bibinfo {author} {\bibfnamefont {R.~C.}\ \bibnamefont
  {{Vermeulen}}}, \bibinfo {author} {\bibfnamefont {P.~M.}\ \bibnamefont
  {{Ogle}}}, \bibinfo {author} {\bibfnamefont {H.~D.}\ \bibnamefont {{Tran}}},
  \bibinfo {author} {\bibfnamefont {I.~W.~A.}\ \bibnamefont {{Browne}}},
  \bibinfo {author} {\bibfnamefont {M.~H.}\ \bibnamefont {{Cohen}}}, \bibinfo
  {author} {\bibfnamefont {A.~C.~S.}\ \bibnamefont {{Readhead}}}, \bibinfo
  {author} {\bibfnamefont {G.~B.}\ \bibnamefont {{Taylor}}}, \ and\ \bibinfo
  {author} {\bibfnamefont {R.~W.}\ \bibnamefont {{Goodrich}}},\ }\href
  {\doibase 10.1086/309716} {\bibfield  {journal} {\bibinfo  {journal} {\apjl}\
  }\textbf {\bibinfo {volume} {452}},\ \bibinfo {pages} {L5} (\bibinfo {year}
  {1995})}\BibitemShut {NoStop}%
\bibitem [{\citenamefont {{Corbett}}\ \emph {et~al.}(2000)\citenamefont
  {{Corbett}}, \citenamefont {{Robinson}}, \citenamefont {{Axon}},\ and\
  \citenamefont {{Hough}}}]{Corbett00}%
  \BibitemOpen
  \bibfield  {author} {\bibinfo {author} {\bibfnamefont {E.~A.}\ \bibnamefont
  {{Corbett}}}, \bibinfo {author} {\bibfnamefont {A.}~\bibnamefont
  {{Robinson}}}, \bibinfo {author} {\bibfnamefont {D.~J.}\ \bibnamefont
  {{Axon}}}, \ and\ \bibinfo {author} {\bibfnamefont {J.~H.}\ \bibnamefont
  {{Hough}}},\ }\href {\doibase 10.1046/j.1365-8711.2000.03045.x} {\bibfield
  {journal} {\bibinfo  {journal} {\mnras}\ }\textbf {\bibinfo {volume} {311}},\
  \bibinfo {pages} {485} (\bibinfo {year} {2000})}\BibitemShut {NoStop}%
\bibitem [{\citenamefont {{Sbarufatti}}\ \emph {et~al.}(2006)\citenamefont
  {{Sbarufatti}}, \citenamefont {{Falomo}}, \citenamefont {{Treves}},\ and\
  \citenamefont {{Kotilainen}}}]{Sbarufatti06}%
  \BibitemOpen
  \bibfield  {author} {\bibinfo {author} {\bibfnamefont {B.}~\bibnamefont
  {{Sbarufatti}}}, \bibinfo {author} {\bibfnamefont {R.}~\bibnamefont
  {{Falomo}}}, \bibinfo {author} {\bibfnamefont {A.}~\bibnamefont {{Treves}}},
  \ and\ \bibinfo {author} {\bibfnamefont {J.}~\bibnamefont {{Kotilainen}}},\
  }\href {\doibase 10.1051/0004-6361:20065455} {\bibfield  {journal} {\bibinfo
  {journal} {\aap}\ }\textbf {\bibinfo {volume} {457}},\ \bibinfo {pages} {35}
  (\bibinfo {year} {2006})},\ \Eprint {http://arxiv.org/abs/astro-ph/0605448}
  {astro-ph/0605448} \BibitemShut {NoStop}%
\bibitem [{\citenamefont {{Capetti}}\ \emph {et~al.}(2010)\citenamefont
  {{Capetti}}, \citenamefont {{Raiteri}},\ and\ \citenamefont
  {{Buttiglione}}}]{Capetti10}%
  \BibitemOpen
  \bibfield  {author} {\bibinfo {author} {\bibfnamefont {A.}~\bibnamefont
  {{Capetti}}}, \bibinfo {author} {\bibfnamefont {C.~M.}\ \bibnamefont
  {{Raiteri}}}, \ and\ \bibinfo {author} {\bibfnamefont {S.}~\bibnamefont
  {{Buttiglione}}},\ }\href {\doibase 10.1051/0004-6361/201014232} {\bibfield
  {journal} {\bibinfo  {journal} {\aap}\ }\textbf {\bibinfo {volume} {516}},\
  \bibinfo {eid} {A59} (\bibinfo {year} {2010})},\ \Eprint
  {http://arxiv.org/abs/1004.2161} {arXiv:1004.2161} \BibitemShut {NoStop}%
\bibitem [{\citenamefont {{Nilsson}}\ \emph {et~al.}(2010)\citenamefont
  {{Nilsson}}, \citenamefont {{Takalo}}, \citenamefont {{Lehto}},\ and\
  \citenamefont {{Sillanp{\"a}{\"a}}}}]{Nilsson10}%
  \BibitemOpen
  \bibfield  {author} {\bibinfo {author} {\bibfnamefont {K.}~\bibnamefont
  {{Nilsson}}}, \bibinfo {author} {\bibfnamefont {L.~O.}\ \bibnamefont
  {{Takalo}}}, \bibinfo {author} {\bibfnamefont {H.~J.}\ \bibnamefont
  {{Lehto}}}, \ and\ \bibinfo {author} {\bibfnamefont {A.}~\bibnamefont
  {{Sillanp{\"a}{\"a}}}},\ }\href {\doibase 10.1051/0004-6361/201014198}
  {\bibfield  {journal} {\bibinfo  {journal} {\aap}\ }\textbf {\bibinfo
  {volume} {516}},\ \bibinfo {eid} {A60} (\bibinfo {year} {2010})},\ \Eprint
  {http://arxiv.org/abs/1004.2617} {arXiv:1004.2617} \BibitemShut {NoStop}%
\bibitem [{\citenamefont {{Landoni}}\ \emph {et~al.}(2012)\citenamefont
  {{Landoni}}, \citenamefont {{Falomo}}, \citenamefont {{Treves}},
  \citenamefont {{Sbarufatti}}, \citenamefont {{Decarli}}, \citenamefont
  {{Tavecchio}},\ and\ \citenamefont {{Kotilainen}}}]{Landoni12}%
  \BibitemOpen
  \bibfield  {author} {\bibinfo {author} {\bibfnamefont {M.}~\bibnamefont
  {{Landoni}}}, \bibinfo {author} {\bibfnamefont {R.}~\bibnamefont {{Falomo}}},
  \bibinfo {author} {\bibfnamefont {A.}~\bibnamefont {{Treves}}}, \bibinfo
  {author} {\bibfnamefont {B.}~\bibnamefont {{Sbarufatti}}}, \bibinfo {author}
  {\bibfnamefont {R.}~\bibnamefont {{Decarli}}}, \bibinfo {author}
  {\bibfnamefont {F.}~\bibnamefont {{Tavecchio}}}, \ and\ \bibinfo {author}
  {\bibfnamefont {J.}~\bibnamefont {{Kotilainen}}},\ }\href {\doibase
  10.1051/0004-6361/201219114} {\bibfield  {journal} {\bibinfo  {journal}
  {\aap}\ }\textbf {\bibinfo {volume} {543}},\ \bibinfo {eid} {A116} (\bibinfo
  {year} {2012})},\ \Eprint {http://arxiv.org/abs/1205.1215} {arXiv:1205.1215}
  \BibitemShut {NoStop}%
\bibitem [{\citenamefont {{Landoni}}\ \emph {et~al.}(2018)\citenamefont
  {{Landoni}}, \citenamefont {{Paiano}}, \citenamefont {{Falomo}},
  \citenamefont {{Scarpa}},\ and\ \citenamefont {{Treves}}}]{Landoni18}%
  \BibitemOpen
  \bibfield  {author} {\bibinfo {author} {\bibfnamefont {M.}~\bibnamefont
  {{Landoni}}}, \bibinfo {author} {\bibfnamefont {S.}~\bibnamefont {{Paiano}}},
  \bibinfo {author} {\bibfnamefont {R.}~\bibnamefont {{Falomo}}}, \bibinfo
  {author} {\bibfnamefont {R.}~\bibnamefont {{Scarpa}}}, \ and\ \bibinfo
  {author} {\bibfnamefont {A.}~\bibnamefont {{Treves}}},\ }\href {\doibase
  10.3847/1538-4357/aac77c} {\bibfield  {journal} {\bibinfo  {journal} {\apj}\
  }\textbf {\bibinfo {volume} {861}},\ \bibinfo {eid} {130} (\bibinfo {year}
  {2018})},\ \Eprint {http://arxiv.org/abs/1805.03888} {arXiv:1805.03888
  [astro-ph.HE]} \BibitemShut {NoStop}%
\bibitem [{\citenamefont {{Dar}}\ and\ \citenamefont {{Laor}}(1997)}]{Dar97}%
  \BibitemOpen
  \bibfield  {author} {\bibinfo {author} {\bibfnamefont {A.}~\bibnamefont
  {{Dar}}}\ and\ \bibinfo {author} {\bibfnamefont {A.}~\bibnamefont {{Laor}}},\
  }\href {\doibase 10.1086/310544} {\bibfield  {journal} {\bibinfo  {journal}
  {\apjl}\ }\textbf {\bibinfo {volume} {478}},\ \bibinfo {pages} {L5} (\bibinfo
  {year} {1997})},\ \Eprint {http://arxiv.org/abs/astro-ph/9610252}
  {astro-ph/9610252} \BibitemShut {NoStop}%
\bibitem [{\citenamefont {{Araudo}}\ \emph {et~al.}(2010)\citenamefont
  {{Araudo}}, \citenamefont {{Bosch-Ramon}},\ and\ \citenamefont
  {{Romero}}}]{Araudo10}%
  \BibitemOpen
  \bibfield  {author} {\bibinfo {author} {\bibfnamefont {A.~T.}\ \bibnamefont
  {{Araudo}}}, \bibinfo {author} {\bibfnamefont {V.}~\bibnamefont
  {{Bosch-Ramon}}}, \ and\ \bibinfo {author} {\bibfnamefont {G.~E.}\
  \bibnamefont {{Romero}}},\ }\href {\doibase 10.1051/0004-6361/201014660}
  {\bibfield  {journal} {\bibinfo  {journal} {\aap}\ }\textbf {\bibinfo
  {volume} {522}},\ \bibinfo {eid} {A97} (\bibinfo {year} {2010})},\ \Eprint
  {http://arxiv.org/abs/1007.2199} {arXiv:1007.2199 [astro-ph.HE]} \BibitemShut
  {NoStop}%
\bibitem [{\citenamefont {{Peterson}}(2006)}]{Peterson06}%
  \BibitemOpen
  \bibfield  {author} {\bibinfo {author} {\bibfnamefont {B.~M.}\ \bibnamefont
  {{Peterson}}},\ }in\ \href {\doibase 10.1007/3-540-34621-X_3} {\emph
  {\bibinfo {booktitle} {Physics of Active Galactic Nuclei at all Scales}}},\
  \bibinfo {series} {Lecture Notes in Physics, Berlin Springer Verlag}, Vol.\
  \bibinfo {volume} {693},\ \bibinfo {editor} {edited by\ \bibinfo {editor}
  {\bibfnamefont {D.}~\bibnamefont {{Alloin}}}}\ (\bibinfo {year} {2006})\
  p.~\bibinfo {pages} {77}\BibitemShut {NoStop}%
\bibitem [{\citenamefont {{Netzer}}(2015)}]{Netzer15}%
  \BibitemOpen
  \bibfield  {author} {\bibinfo {author} {\bibfnamefont {H.}~\bibnamefont
  {{Netzer}}},\ }\href {\doibase 10.1146/annurev-astro-082214-122302}
  {\bibfield  {journal} {\bibinfo  {journal} {\araa}\ }\textbf {\bibinfo
  {volume} {53}},\ \bibinfo {pages} {365} (\bibinfo {year} {2015})},\ \Eprint
  {http://arxiv.org/abs/1505.00811} {arXiv:1505.00811} \BibitemShut {NoStop}%
\bibitem [{\citenamefont {{Baldwin}}\ \emph {et~al.}(2003)\citenamefont
  {{Baldwin}}, \citenamefont {{Ferland}}, \citenamefont {{Korista}},
  \citenamefont {{Hamann}},\ and\ \citenamefont {{Dietrich}}}]{Baldwin03}%
  \BibitemOpen
  \bibfield  {author} {\bibinfo {author} {\bibfnamefont {J.~A.}\ \bibnamefont
  {{Baldwin}}}, \bibinfo {author} {\bibfnamefont {G.~J.}\ \bibnamefont
  {{Ferland}}}, \bibinfo {author} {\bibfnamefont {K.~T.}\ \bibnamefont
  {{Korista}}}, \bibinfo {author} {\bibfnamefont {F.}~\bibnamefont {{Hamann}}},
  \ and\ \bibinfo {author} {\bibfnamefont {M.}~\bibnamefont {{Dietrich}}},\
  }\href {\doibase 10.1086/344788} {\bibfield  {journal} {\bibinfo  {journal}
  {\apj}\ }\textbf {\bibinfo {volume} {582}},\ \bibinfo {pages} {590} (\bibinfo
  {year} {2003})},\ \Eprint {http://arxiv.org/abs/astro-ph/0209335}
  {astro-ph/0209335} \BibitemShut {NoStop}%
\bibitem [{\citenamefont {{Zhang}}\ and\ \citenamefont
  {{Yan}}(2011)}]{Zhang11}%
  \BibitemOpen
  \bibfield  {author} {\bibinfo {author} {\bibfnamefont {B.}~\bibnamefont
  {{Zhang}}}\ and\ \bibinfo {author} {\bibfnamefont {H.}~\bibnamefont
  {{Yan}}},\ }\href {\doibase 10.1088/0004-637X/726/2/90} {\bibfield  {journal}
  {\bibinfo  {journal} {\apj}\ }\textbf {\bibinfo {volume} {726}},\ \bibinfo
  {eid} {90} (\bibinfo {year} {2011})},\ \Eprint
  {http://arxiv.org/abs/1011.1197} {arXiv:1011.1197 [astro-ph.HE]} \BibitemShut
  {NoStop}%
\bibitem [{\citenamefont {{Stern}}\ and\ \citenamefont
  {{Laor}}(2012)}]{Stern12}%
  \BibitemOpen
  \bibfield  {author} {\bibinfo {author} {\bibfnamefont {J.}~\bibnamefont
  {{Stern}}}\ and\ \bibinfo {author} {\bibfnamefont {A.}~\bibnamefont
  {{Laor}}},\ }\href {\doibase 10.1111/j.1365-2966.2012.21772.x} {\bibfield
  {journal} {\bibinfo  {journal} {\mnras}\ }\textbf {\bibinfo {volume} {426}},\
  \bibinfo {pages} {2703} (\bibinfo {year} {2012})},\ \Eprint
  {http://arxiv.org/abs/1207.5543} {arXiv:1207.5543} \BibitemShut {NoStop}%
\bibitem [{\citenamefont {{Dermer}}\ \emph {et~al.}(2012)\citenamefont
  {{Dermer}}, \citenamefont {{Murase}},\ and\ \citenamefont
  {{Takami}}}]{Dermer12}%
  \BibitemOpen
  \bibfield  {author} {\bibinfo {author} {\bibfnamefont {C.~D.}\ \bibnamefont
  {{Dermer}}}, \bibinfo {author} {\bibfnamefont {K.}~\bibnamefont {{Murase}}},
  \ and\ \bibinfo {author} {\bibfnamefont {H.}~\bibnamefont {{Takami}}},\
  }\href {\doibase 10.1088/0004-637X/755/2/147} {\bibfield  {journal} {\bibinfo
   {journal} {\apj}\ }\textbf {\bibinfo {volume} {755}},\ \bibinfo {eid} {147}
  (\bibinfo {year} {2012})},\ \Eprint {http://arxiv.org/abs/1203.6544}
  {arXiv:1203.6544 [astro-ph.HE]} \BibitemShut {NoStop}%
\bibitem [{\citenamefont {{Murase}}\ \emph {et~al.}(2016)\citenamefont
  {{Murase}}, \citenamefont {{Guetta}},\ and\ \citenamefont
  {{Ahlers}}}]{Murase16}%
  \BibitemOpen
  \bibfield  {author} {\bibinfo {author} {\bibfnamefont {K.}~\bibnamefont
  {{Murase}}}, \bibinfo {author} {\bibfnamefont {D.}~\bibnamefont {{Guetta}}},
  \ and\ \bibinfo {author} {\bibfnamefont {M.}~\bibnamefont {{Ahlers}}},\
  }\href {\doibase 10.1103/PhysRevLett.116.071101} {\bibfield  {journal}
  {\bibinfo  {journal} {Physical Review Letters}\ }\textbf {\bibinfo {volume}
  {116}},\ \bibinfo {eid} {071101} (\bibinfo {year} {2016})},\ \Eprint
  {http://arxiv.org/abs/1509.00805} {arXiv:1509.00805 [astro-ph.HE]}
  \BibitemShut {NoStop}%
\bibitem [{\citenamefont {{Inoue}}\ and\ \citenamefont
  {{Takahara}}(1996)}]{Inoue96}%
  \BibitemOpen
  \bibfield  {author} {\bibinfo {author} {\bibfnamefont {S.}~\bibnamefont
  {{Inoue}}}\ and\ \bibinfo {author} {\bibfnamefont {F.}~\bibnamefont
  {{Takahara}}},\ }\href {\doibase 10.1086/177270} {\bibfield  {journal}
  {\bibinfo  {journal} {\apj}\ }\textbf {\bibinfo {volume} {463}},\ \bibinfo
  {pages} {555} (\bibinfo {year} {1996})}\BibitemShut {NoStop}%
\bibitem [{\citenamefont {{Tavecchio}}\ \emph {et~al.}(1998)\citenamefont
  {{Tavecchio}}, \citenamefont {{Maraschi}},\ and\ \citenamefont
  {{Ghisellini}}}]{Tavecchio98}%
  \BibitemOpen
  \bibfield  {author} {\bibinfo {author} {\bibfnamefont {F.}~\bibnamefont
  {{Tavecchio}}}, \bibinfo {author} {\bibfnamefont {L.}~\bibnamefont
  {{Maraschi}}}, \ and\ \bibinfo {author} {\bibfnamefont {G.}~\bibnamefont
  {{Ghisellini}}},\ }\href {\doibase 10.1086/306526} {\bibfield  {journal}
  {\bibinfo  {journal} {\apj}\ }\textbf {\bibinfo {volume} {509}},\ \bibinfo
  {pages} {608} (\bibinfo {year} {1998})},\ \Eprint
  {http://arxiv.org/abs/astro-ph/9809051} {astro-ph/9809051} \BibitemShut
  {NoStop}%
\bibitem [{\citenamefont {{Costamante}}\ and\ \citenamefont
  {{Ghisellini}}(2002)}]{Costamante02}%
  \BibitemOpen
  \bibfield  {author} {\bibinfo {author} {\bibfnamefont {L.}~\bibnamefont
  {{Costamante}}}\ and\ \bibinfo {author} {\bibfnamefont {G.}~\bibnamefont
  {{Ghisellini}}},\ }\href {\doibase 10.1051/0004-6361:20011749} {\bibfield
  {journal} {\bibinfo  {journal} {\aap}\ }\textbf {\bibinfo {volume} {384}},\
  \bibinfo {pages} {56} (\bibinfo {year} {2002})},\ \Eprint
  {http://arxiv.org/abs/astro-ph/0112201} {astro-ph/0112201} \BibitemShut
  {NoStop}%
\bibitem [{\citenamefont {{Chartas}}\ \emph {et~al.}(2000)\citenamefont
  {{Chartas}}, \citenamefont {{Worrall}}, \citenamefont {{Birkinshaw}},
  \citenamefont {{Cresitello-Dittmar}}, \citenamefont {{Cui}}, \citenamefont
  {{Ghosh}}, \citenamefont {{Harris}}, \citenamefont {{Hooper}}, \citenamefont
  {{Jauncey}}, \citenamefont {{Kim}}, \citenamefont {{Lovell}}, \citenamefont
  {{Mathur}}, \citenamefont {{Schwartz}}, \citenamefont {{Tingay}},
  \citenamefont {{Virani}},\ and\ \citenamefont {{Wilkes}}}]{Chartas00}%
  \BibitemOpen
  \bibfield  {author} {\bibinfo {author} {\bibfnamefont {G.}~\bibnamefont
  {{Chartas}}}, \bibinfo {author} {\bibfnamefont {D.~M.}\ \bibnamefont
  {{Worrall}}}, \bibinfo {author} {\bibfnamefont {M.}~\bibnamefont
  {{Birkinshaw}}}, \bibinfo {author} {\bibfnamefont {M.}~\bibnamefont
  {{Cresitello-Dittmar}}}, \bibinfo {author} {\bibfnamefont {W.}~\bibnamefont
  {{Cui}}}, \bibinfo {author} {\bibfnamefont {K.~K.}\ \bibnamefont {{Ghosh}}},
  \bibinfo {author} {\bibfnamefont {D.~E.}\ \bibnamefont {{Harris}}}, \bibinfo
  {author} {\bibfnamefont {E.~J.}\ \bibnamefont {{Hooper}}}, \bibinfo {author}
  {\bibfnamefont {D.~L.}\ \bibnamefont {{Jauncey}}}, \bibinfo {author}
  {\bibfnamefont {D.-W.}\ \bibnamefont {{Kim}}}, \bibinfo {author}
  {\bibfnamefont {J.}~\bibnamefont {{Lovell}}}, \bibinfo {author}
  {\bibfnamefont {S.}~\bibnamefont {{Mathur}}}, \bibinfo {author}
  {\bibfnamefont {D.~A.}\ \bibnamefont {{Schwartz}}}, \bibinfo {author}
  {\bibfnamefont {S.~J.}\ \bibnamefont {{Tingay}}}, \bibinfo {author}
  {\bibfnamefont {S.~N.}\ \bibnamefont {{Virani}}}, \ and\ \bibinfo {author}
  {\bibfnamefont {B.~J.}\ \bibnamefont {{Wilkes}}},\ }\href {\doibase
  10.1086/317049} {\bibfield  {journal} {\bibinfo  {journal} {\apj}\ }\textbf
  {\bibinfo {volume} {542}},\ \bibinfo {pages} {655} (\bibinfo {year}
  {2000})},\ \Eprint {http://arxiv.org/abs/astro-ph/0005227} {astro-ph/0005227}
  \BibitemShut {NoStop}%
\bibitem [{\citenamefont {{Jester}}\ \emph {et~al.}(2006)\citenamefont
  {{Jester}}, \citenamefont {{Harris}}, \citenamefont {{Marshall}},\ and\
  \citenamefont {{Meisenheimer}}}]{Jester06}%
  \BibitemOpen
  \bibfield  {author} {\bibinfo {author} {\bibfnamefont {S.}~\bibnamefont
  {{Jester}}}, \bibinfo {author} {\bibfnamefont {D.~E.}\ \bibnamefont
  {{Harris}}}, \bibinfo {author} {\bibfnamefont {H.~L.}\ \bibnamefont
  {{Marshall}}}, \ and\ \bibinfo {author} {\bibfnamefont {K.}~\bibnamefont
  {{Meisenheimer}}},\ }\href {\doibase 10.1086/505962} {\bibfield  {journal}
  {\bibinfo  {journal} {\apj}\ }\textbf {\bibinfo {volume} {648}},\ \bibinfo
  {pages} {900} (\bibinfo {year} {2006})},\ \Eprint
  {http://arxiv.org/abs/astro-ph/0605529} {astro-ph/0605529} \BibitemShut
  {NoStop}%
\bibitem [{\citenamefont {{Kelner}}\ \emph {et~al.}(2006)\citenamefont
  {{Kelner}}, \citenamefont {{Aharonian}},\ and\ \citenamefont
  {{Bugayov}}}]{Kelner06}%
  \BibitemOpen
  \bibfield  {author} {\bibinfo {author} {\bibfnamefont {S.~R.}\ \bibnamefont
  {{Kelner}}}, \bibinfo {author} {\bibfnamefont {F.~A.}\ \bibnamefont
  {{Aharonian}}}, \ and\ \bibinfo {author} {\bibfnamefont {V.~V.}\ \bibnamefont
  {{Bugayov}}},\ }\href {\doibase 10.1103/PhysRevD.74.034018} {\bibfield
  {journal} {\bibinfo  {journal} {\prd}\ }\textbf {\bibinfo {volume} {74}},\
  \bibinfo {eid} {034018} (\bibinfo {year} {2006})},\ \Eprint
  {http://arxiv.org/abs/astro-ph/0606058} {astro-ph/0606058} \BibitemShut
  {NoStop}%
\bibitem [{\citenamefont {{Stecker}}(1970)}]{Stecker70}%
  \BibitemOpen
  \bibfield  {author} {\bibinfo {author} {\bibfnamefont {F.~W.}\ \bibnamefont
  {{Stecker}}},\ }\href {\doibase 10.1007/BF00653856} {\bibfield  {journal}
  {\bibinfo  {journal} {\apss}\ }\textbf {\bibinfo {volume} {6}},\ \bibinfo
  {pages} {377} (\bibinfo {year} {1970})}\BibitemShut {NoStop}%
\bibitem [{\citenamefont {{Kafexhiu}}\ \emph {et~al.}(2014)\citenamefont
  {{Kafexhiu}}, \citenamefont {{Aharonian}}, \citenamefont {{Taylor}},\ and\
  \citenamefont {{Vila}}}]{Ervin14}%
  \BibitemOpen
  \bibfield  {author} {\bibinfo {author} {\bibfnamefont {E.}~\bibnamefont
  {{Kafexhiu}}}, \bibinfo {author} {\bibfnamefont {F.}~\bibnamefont
  {{Aharonian}}}, \bibinfo {author} {\bibfnamefont {A.~M.}\ \bibnamefont
  {{Taylor}}}, \ and\ \bibinfo {author} {\bibfnamefont {G.~S.}\ \bibnamefont
  {{Vila}}},\ }\href {\doibase 10.1103/PhysRevD.90.123014} {\bibfield
  {journal} {\bibinfo  {journal} {\prd}\ }\textbf {\bibinfo {volume} {90}},\
  \bibinfo {eid} {123014} (\bibinfo {year} {2014})},\ \Eprint
  {http://arxiv.org/abs/1406.7369} {arXiv:1406.7369 [astro-ph.HE]} \BibitemShut
  {NoStop}%
\bibitem [{\citenamefont {{B{\"o}ttcher}}\ \emph {et~al.}(2013)\citenamefont
  {{B{\"o}ttcher}}, \citenamefont {{Reimer}}, \citenamefont {{Sweeney}},\ and\
  \citenamefont {{Prakash}}}]{Boettcher13}%
  \BibitemOpen
  \bibfield  {author} {\bibinfo {author} {\bibfnamefont {M.}~\bibnamefont
  {{B{\"o}ttcher}}}, \bibinfo {author} {\bibfnamefont {A.}~\bibnamefont
  {{Reimer}}}, \bibinfo {author} {\bibfnamefont {K.}~\bibnamefont {{Sweeney}}},
  \ and\ \bibinfo {author} {\bibfnamefont {A.}~\bibnamefont {{Prakash}}},\
  }\href {\doibase 10.1088/0004-637X/768/1/54} {\bibfield  {journal} {\bibinfo
  {journal} {\apj}\ }\textbf {\bibinfo {volume} {768}},\ \bibinfo {eid} {54}
  (\bibinfo {year} {2013})},\ \Eprint {http://arxiv.org/abs/1304.0605}
  {arXiv:1304.0605 [astro-ph.HE]} \BibitemShut {NoStop}%
\bibitem [{\citenamefont {{Wang}}\ \emph {et~al.}(2018)\citenamefont {{Wang}},
  \citenamefont {{Liu}}, \citenamefont {{Dai}},\ and\ \citenamefont
  {{Asano}}}]{Wang18}%
  \BibitemOpen
  \bibfield  {author} {\bibinfo {author} {\bibfnamefont {K.}~\bibnamefont
  {{Wang}}}, \bibinfo {author} {\bibfnamefont {R.-Y.}\ \bibnamefont {{Liu}}},
  \bibinfo {author} {\bibfnamefont {Z.-G.}\ \bibnamefont {{Dai}}}, \ and\
  \bibinfo {author} {\bibfnamefont {K.}~\bibnamefont {{Asano}}},\ }\href
  {\doibase 10.3847/1538-4357/aab667} {\bibfield  {journal} {\bibinfo
  {journal} {\apj}\ }\textbf {\bibinfo {volume} {857}},\ \bibinfo {eid} {24}
  (\bibinfo {year} {2018})},\ \Eprint {http://arxiv.org/abs/1803.04112}
  {arXiv:1803.04112 [astro-ph.HE]} \BibitemShut {NoStop}%
\bibitem [{\citenamefont {{Draine}}(2011)}]{Draine11}%
  \BibitemOpen
  \bibfield  {author} {\bibinfo {author} {\bibfnamefont {B.~T.}\ \bibnamefont
  {{Draine}}},\ }\href@noop {} {\emph {\bibinfo {title} {Physics of the
  Interstellar and Intergalactic Medium by Bruce T.~Draine.~Princeton
  University Press, 2011.~ISBN: 978-0-691-12214-4}}}\ (\bibinfo {year}
  {2011})\BibitemShut {NoStop}%
\bibitem [{\citenamefont {{Finke}}\ \emph {et~al.}(2010)\citenamefont
  {{Finke}}, \citenamefont {{Razzaque}},\ and\ \citenamefont
  {{Dermer}}}]{Finke10}%
  \BibitemOpen
  \bibfield  {author} {\bibinfo {author} {\bibfnamefont {J.~D.}\ \bibnamefont
  {{Finke}}}, \bibinfo {author} {\bibfnamefont {S.}~\bibnamefont {{Razzaque}}},
  \ and\ \bibinfo {author} {\bibfnamefont {C.~D.}\ \bibnamefont {{Dermer}}},\
  }\href {\doibase 10.1088/0004-637X/712/1/238} {\bibfield  {journal} {\bibinfo
   {journal} {\apj}\ }\textbf {\bibinfo {volume} {712}},\ \bibinfo {pages}
  {238} (\bibinfo {year} {2010})},\ \Eprint {http://arxiv.org/abs/0905.1115}
  {arXiv:0905.1115 [astro-ph.HE]} \BibitemShut {NoStop}%
\bibitem [{\citenamefont {{Aleksi{\'c}}}\ \emph {et~al.}(2014)\citenamefont
  {{Aleksi{\'c}}}, \citenamefont {{Ansoldi}}, \citenamefont {{Antonelli}},
  \citenamefont {{Antoranz}}, \citenamefont {{Babic}},\ and\ \citenamefont
  {et~al.}}]{Magic14}%
  \BibitemOpen
  \bibfield  {author} {\bibinfo {author} {\bibfnamefont {J.}~\bibnamefont
  {{Aleksi{\'c}}}}, \bibinfo {author} {\bibfnamefont {S.}~\bibnamefont
  {{Ansoldi}}}, \bibinfo {author} {\bibfnamefont {L.~A.}\ \bibnamefont
  {{Antonelli}}}, \bibinfo {author} {\bibfnamefont {P.}~\bibnamefont
  {{Antoranz}}}, \bibinfo {author} {\bibfnamefont {A.}~\bibnamefont {{Babic}}},
  \ and\ \bibinfo {author} {\bibnamefont {et~al.}},\ }\href {\doibase
  10.1051/0004-6361/201423364} {\bibfield  {journal} {\bibinfo  {journal}
  {\aap}\ }\textbf {\bibinfo {volume} {567}},\ \bibinfo {eid} {A135} (\bibinfo
  {year} {2014})},\ \Eprint {http://arxiv.org/abs/1401.0464} {arXiv:1401.0464
  [astro-ph.HE]} \BibitemShut {NoStop}%
\bibitem [{\citenamefont {{Gao}}\ \emph {et~al.}(2017)\citenamefont {{Gao}},
  \citenamefont {{Pohl}},\ and\ \citenamefont {{Winter}}}]{Gao17}%
  \BibitemOpen
  \bibfield  {author} {\bibinfo {author} {\bibfnamefont {S.}~\bibnamefont
  {{Gao}}}, \bibinfo {author} {\bibfnamefont {M.}~\bibnamefont {{Pohl}}}, \
  and\ \bibinfo {author} {\bibfnamefont {W.}~\bibnamefont {{Winter}}},\ }\href
  {\doibase 10.3847/1538-4357/aa7754} {\bibfield  {journal} {\bibinfo
  {journal} {\apj}\ }\textbf {\bibinfo {volume} {843}},\ \bibinfo {eid} {109}
  (\bibinfo {year} {2017})},\ \Eprint {http://arxiv.org/abs/1610.05306}
  {arXiv:1610.05306 [astro-ph.HE]} \BibitemShut {NoStop}%
\bibitem [{\citenamefont {{Stecker}}\ \emph {et~al.}(1991)\citenamefont
  {{Stecker}}, \citenamefont {{Done}}, \citenamefont {{Salamon}},\ and\
  \citenamefont {{Sommers}}}]{Stecker91}%
  \BibitemOpen
  \bibfield  {author} {\bibinfo {author} {\bibfnamefont {F.~W.}\ \bibnamefont
  {{Stecker}}}, \bibinfo {author} {\bibfnamefont {C.}~\bibnamefont {{Done}}},
  \bibinfo {author} {\bibfnamefont {M.~H.}\ \bibnamefont {{Salamon}}}, \ and\
  \bibinfo {author} {\bibfnamefont {P.}~\bibnamefont {{Sommers}}},\ }\href
  {\doibase 10.1103/PhysRevLett.66.2697} {\bibfield  {journal} {\bibinfo
  {journal} {Physical Review Letters}\ }\textbf {\bibinfo {volume} {66}},\
  \bibinfo {pages} {2697} (\bibinfo {year} {1991})}\BibitemShut {NoStop}%
\bibitem [{\citenamefont {{Mannheim}}(1993)}]{Mannheim93}%
  \BibitemOpen
  \bibfield  {author} {\bibinfo {author} {\bibfnamefont {K.}~\bibnamefont
  {{Mannheim}}},\ }\href {\doibase 10.1103/PhysRevD.48.2408} {\bibfield
  {journal} {\bibinfo  {journal} {\prd}\ }\textbf {\bibinfo {volume} {48}},\
  \bibinfo {pages} {2408} (\bibinfo {year} {1993})},\ \Eprint
  {http://arxiv.org/abs/astro-ph/9306005} {astro-ph/9306005} \BibitemShut
  {NoStop}%
\bibitem [{\citenamefont {{Atoyan}}\ and\ \citenamefont
  {{Dermer}}(2001)}]{Atoyan01}%
  \BibitemOpen
  \bibfield  {author} {\bibinfo {author} {\bibfnamefont {A.}~\bibnamefont
  {{Atoyan}}}\ and\ \bibinfo {author} {\bibfnamefont {C.~D.}\ \bibnamefont
  {{Dermer}}},\ }\href {\doibase 10.1103/PhysRevLett.87.221102} {\bibfield
  {journal} {\bibinfo  {journal} {Physical Review Letters}\ }\textbf {\bibinfo
  {volume} {87}},\ \bibinfo {eid} {221102} (\bibinfo {year} {2001})},\ \Eprint
  {http://arxiv.org/abs/astro-ph/0108053} {astro-ph/0108053} \BibitemShut
  {NoStop}%
\bibitem [{\citenamefont {{M{\"u}cke}}\ \emph {et~al.}(2003)\citenamefont
  {{M{\"u}cke}}, \citenamefont {{Protheroe}}, \citenamefont {{Engel}},
  \citenamefont {{Rachen}},\ and\ \citenamefont {{Stanev}}}]{Mucke03}%
  \BibitemOpen
  \bibfield  {author} {\bibinfo {author} {\bibfnamefont {A.}~\bibnamefont
  {{M{\"u}cke}}}, \bibinfo {author} {\bibfnamefont {R.~J.}\ \bibnamefont
  {{Protheroe}}}, \bibinfo {author} {\bibfnamefont {R.}~\bibnamefont
  {{Engel}}}, \bibinfo {author} {\bibfnamefont {J.~P.}\ \bibnamefont
  {{Rachen}}}, \ and\ \bibinfo {author} {\bibfnamefont {T.}~\bibnamefont
  {{Stanev}}},\ }\href {\doibase 10.1016/S0927-6505(02)00185-8} {\bibfield
  {journal} {\bibinfo  {journal} {Astroparticle Physics}\ }\textbf {\bibinfo
  {volume} {18}},\ \bibinfo {pages} {593} (\bibinfo {year} {2003})},\ \Eprint
  {http://arxiv.org/abs/astro-ph/0206164} {astro-ph/0206164} \BibitemShut
  {NoStop}%
\bibitem [{\citenamefont {{IceCube Collaboration}}(2018)}]{IC18_TXSnuflare}%
  \BibitemOpen
  \bibfield  {author} {\bibinfo {author} {\bibnamefont {{IceCube
  Collaboration}}},\ }\href {\doibase 10.1126/science.aat2890} {\bibfield
  {journal} {\bibinfo  {journal} {Science}\ }\textbf {\bibinfo {volume}
  {361}},\ \bibinfo {pages} {147} (\bibinfo {year} {2018})}\BibitemShut
  {NoStop}%
\bibitem [{\citenamefont {{Pittard}}\ \emph {et~al.}(2010)\citenamefont
  {{Pittard}}, \citenamefont {{Hartquist}},\ and\ \citenamefont
  {{Falle}}}]{Pittard10}%
  \BibitemOpen
  \bibfield  {author} {\bibinfo {author} {\bibfnamefont {J.~M.}\ \bibnamefont
  {{Pittard}}}, \bibinfo {author} {\bibfnamefont {T.~W.}\ \bibnamefont
  {{Hartquist}}}, \ and\ \bibinfo {author} {\bibfnamefont {S.~A.~E.~G.}\
  \bibnamefont {{Falle}}},\ }\href {\doibase 10.1111/j.1365-2966.2010.16504.x}
  {\bibfield  {journal} {\bibinfo  {journal} {\mnras}\ }\textbf {\bibinfo
  {volume} {405}},\ \bibinfo {pages} {821} (\bibinfo {year} {2010})},\ \Eprint
  {http://arxiv.org/abs/1002.2091} {arXiv:1002.2091} \BibitemShut {NoStop}%
\bibitem [{\citenamefont {{Ginzburg}}\ and\ \citenamefont
  {{Syrovatskii}}(1964)}]{Ginzburg64}%
  \BibitemOpen
  \bibfield  {author} {\bibinfo {author} {\bibfnamefont {V.~L.}\ \bibnamefont
  {{Ginzburg}}}\ and\ \bibinfo {author} {\bibfnamefont {S.~I.}\ \bibnamefont
  {{Syrovatskii}}},\ }\href@noop {} {\emph {\bibinfo {title} {The Origin of
  Cosmic Rays, New York: Macmillan, 1964}}}\ (\bibinfo {year}
  {1964})\BibitemShut {NoStop}%
\bibitem [{\citenamefont {{Aharonian}}\ \emph {et~al.}(1983)\citenamefont
  {{Aharonian}}, \citenamefont {{Atoyan}},\ and\ \citenamefont
  {{Nagapetyan}}}]{Aharonian83_epp}%
  \BibitemOpen
  \bibfield  {author} {\bibinfo {author} {\bibfnamefont {F.~A.}\ \bibnamefont
  {{Aharonian}}}, \bibinfo {author} {\bibfnamefont {A.~M.}\ \bibnamefont
  {{Atoyan}}}, \ and\ \bibinfo {author} {\bibfnamefont {A.~M.}\ \bibnamefont
  {{Nagapetyan}}},\ }\href {\doibase 10.1007/BF01005624} {\bibfield  {journal}
  {\bibinfo  {journal} {Astrophysics}\ }\textbf {\bibinfo {volume} {19}},\
  \bibinfo {pages} {187} (\bibinfo {year} {1983})}\BibitemShut {NoStop}%
\end{thebibliography}%

\end{document}